\documentclass[12pt]{article}
\usepackage[dvipsnames,svgnames]{xcolor}
\usepackage{amsfonts,amssymb,amsmath}
\usepackage{subfigure}
\usepackage{graphicx}
\usepackage{appendix}
\usepackage{euler}
\usepackage{pifont} 
\usepackage{enumerate}
\usepackage{tikz}
\usetikzlibrary{shapes,arrows,shadows,matrix}
\usetikzlibrary{shapes.gates.logic.US,trees,positioning}
\usepackage{fancyhdr}
\usepackage{hyperref}

            \newcommand{\ec}{\varepsilon^2}
            \newcommand{\eps}{\varepsilon}

            \newcommand{\V}{\mathbf{V}}
            \newcommand{\uv}{\mathbf{u}}
            \newcommand{\q}{\mathbf{q}}
            \newcommand{\n}{\mathbf{n}}
            
            \newcommand{\Vt}{\tilde\V}
            
            \newcommand{\minmod}{\text{minmod}}
            \newcommand{\Dx}{\Delta x}
            
            \newcommand{\Dt}{\Delta t}
            \newcommand{\ud}{\frac{1}{2}}

\numberwithin{equation}{section}

\title{An Asymptotic-Preserving all-speed scheme for the Euler and Navier-Stokes equations}

\author{Floraine Cordier$^{1,2,3}$, Pierre Degond$^{2,3}$, Anela Kumbaro$^{1}$ \\     
\small  $^{1}$ CEA-Saclay  \textsc{DEN, DM2S, SFME, LETR}  F-91191 Gif-sur-Yvette, France.\\    
\small floraine.cordier@cea.fr, anela.kumbaro@cea.fr\\
\small  $^{2}$ Universit\'e de Toulouse; UPS, INSA, UT1, UTM ; Institut de Math\'ematiques de Toulouse ; \\
\small  F-31062 Toulouse, France. \\
\small  $^{3}$ CNRS; Institut de Math\'ematiques de Toulouse UMR 5219 ;
F-31062 Toulouse, France.\\
\small pierre.degond@math.univ-toulouse.fr 
}

\date{}

\begin{document}

\maketitle

\vspace{0.1 cm}


\begin{abstract}

We present an Asymptotic-Preserving 'all-speed' scheme for the simulation of compressible flows valid at all Mach-numbers ranging from very small to order unity. The scheme is based on a semi-implicit discretization which treats the acoustic part implicitly and the convective and diffusive parts explicitly. This discretization, which is the key to the Asymptotic-Preserving  property, provides a consistent approximation of both the hyperbolic compressible regime and the elliptic incompressible regime. The divergence-free condition on the velocity in the incompressible regime is respected, and an the pressure is computed via an elliptic equation resulting from a suitable combination of the momentum and energy equations. The implicit treatment of the acoustic part allows the time-step to be independent of the Mach number. The scheme is conservative and applies to steady or unsteady flows and to general equations of state. One and Two-dimensional numerical results provide a validation of the Asymptotic-Preserving 'all-speed' properties.
\end{abstract}





\bigskip

\setcounter{equation}{0}
\section{Introduction}
\label{sec_intro}

The numerical simulation of fluid flows at all Mach numbers is an active field of research. The occurrence of low Mach number regions in a globally compressible flow may be caused by the boundary or initial conditions (e.g. in a fluid at rest subject to a supersonic jet ), by the geometry of the problem (e.g. in a nozzle with a large variation of the section), or by the underlying Physics (e.g. in the case of phase changes). This occurrence gives rise to specific numerical issues which are discussed below. 

When the Mach number tends to zero, compressible flow equations converge to incompressible equations: the compressible Euler equations in the inviscid case (respectively the compressible Navier-Stokes equations in the viscous case) converge to the incompressible Euler equations (respectively incompressible Navier-Stokes equations). This convergence has been studied mathematically by Klainerman and Majda \cite{klainerman1981singular, klainerman1982compressible} (See also \cite{danchin2005low, gallagher2005resultats, metivier2001incompressible, schochet2005mathematical} for reviews and references). However, in numerical simulations, it is very difficult to shift from compressible flow equations to incompressible ones in the regions where the Mach-number becomes very small. Therefore, it is necessary to design numerical methods for compressible flows that can handle both the compressible regime (i.e. local Mach-number of order unity) and the incompressible one (i.e. very small local Mach-number). This is the purpose of \textit{'All-Speed schemes'}.

In this work, we derive an All-Speed scheme using the Asymptotic-Preserving methodology. The Asymptotic-Preserving (AP) property is defined as follows. Consider a continuous physical model ${\cal{M}}^\eps$ which involves a perturbation parameter $\eps$ (here, $\eps$ is the scaled Mach-number and ${\cal{M}}^\eps$ represents the compressible Euler or Navier-Stokes model) which can range from $\eps = {\mathcal O}(1)$ to $\eps \ll 1$ values. Let ${\cal{M}}^{0}$ the limit of ${\cal{M}}^\eps$ when $\eps \to 0$ (here ${\cal{M}}^{0}$ is the incompressible Euler or Navier-Stokes model). Let now ${\cal{M}}^\eps_\Delta$ be a numerical scheme which provides a consistent discretization of ${\cal{M}}^\eps$ with discrete time and space steps $(\Dt, \Delta x)=\Delta$. The scheme ${\cal{M}}^\eps_\Delta$ is said to be \textit{Asymptotic-Preserving (AP)} if its stability condition is independent of $\eps$ and if its limit ${\cal{M}}^0_\Delta$ as $\eps \to 0$ provides a consistent discretization of the continuous limit model ${\cal{M}}^{0}$. The AP property is illustrated by the commutative diagram of fig. \ref{fig:ap}.

\begin{figure}[h!]
\begin{center}
\begin{tikzpicture}[description/.style={fill=white,inner sep=2pt}]
\matrix (m) [matrix of math nodes, row sep=3em,
column sep=4em, text height=3ex, text depth=0.5ex]
{ {\cal{M}}^\eps &  {\cal{M}}^{0} \\
{\cal{M}}^\eps_\Delta &   {\cal{M}}^0_\Delta \\ };

\path[->,font=\scriptsize]
(m-1-1) edge node[auto] {$\eps \to 0 $} (m-1-2)
(m-2-1) edge node[auto] {$ \Delta \to 0 $} (m-1-1);

\path[<-,font=\scriptsize]
(m-1-2) edge node[auto] {$ \Delta \to 0$} (m-2-2)
(m-2-2) edge node[auto] {$ \eps \to 0 $} (m-2-1);
\end{tikzpicture}
\caption{Asymptotic-Preserving (AP) property: the upper horizontal arrow translates the assumption that the continuous model ${\cal{M}}^\eps$ tends to the limit model ${\cal{M}}^{0}$ when $\eps \to 0$. The left vertical arrow expresses that ${\cal{M}}^\eps_\Delta$ is a consistent discretization of ${\cal{M}}^\eps$ when the discretization parameter $\Delta \to 0$. The lower horizontal arrow indicates that the scheme ${\cal{M}}^\eps_\Delta$ has a limit ${\cal{M}}^0_\Delta$ when $\eps \to 0$ for fixed $\Delta$. Finally, the right vertical arrow expresses the AP-property: it says that the limit scheme ${\cal{M}}^0_\Delta$ is a consistent discretization of the limit model ${\cal{M}}^{0}$ when $\Delta \to 0$.}
\label{fig:ap}
\end{center}
\end{figure}

The present scheme is derived following the AP methodology and targets the situation of mixed flows where part of the flow has local Mach-number of order unity and is in a compressible regime and part of the flow has very small local Mach-number and is in the incompressible regime. More precisely, our scheme meets the following requirements. It is AP, i.e. it is consistent with both the compressible and incompressible regimes. The divergence-free condition on the velocity in the incompressible regime is explicitly satisfied up to the order of the approximation. The CFL condition is independent of the Mach-number. Therefore, the time-step is not constrained to be inversely proportional to the sound speed like. We remind that classical explicit schemes require such a time-step constraint which is very detrimental to the scheme efficiency  in the small Mach-number regime. The scheme is conservative and preserves the correct shock speeds in the compressible regime. At last, the scheme applies to a general equation of state and to steady as well as unsteady flows. 

The present work is the continuation of earlier work on the construction of Asymptotic-Preserving schemes for fluid equations in the small Mach-number limit. In \cite{min}, a first-order AP scheme is derived for the isentropic Euler equations. A second order version of this scheme based on the Kurganov-Tadmor central scheme methodology is proposed in \cite{min2}. Here, we extend the work of \cite{min} to the full Euler and Navier-Stokes equations, i.e. including an energy equations instead of the isentropic assumption. This addition involves more than a simple technical adaptation. Indeed, the scheme has to be strongly modified in the choice of the terms that require an implicit treatment. Some of these terms have to be shifted from the mass to the energy conservation equation. With the use of a real gas equation of state, the resulting pressure equation becomes nonlinear and requires a specific treatment. We also provide a second-order extension of the method based on the classical MUSCL methodology which can apply to a larger software framework than the central scheme methodology. The numerical results will show that the passage to second order is qualitatively necessary to achieve a good accuracy. We also mention \cite{haack2010all} which relates to \cite{min} but provides an alternate way of reaching the AP-property. 

Understanding why compressible flow solvers perform so poorly in the low Mach-number regime has triggered a vast literature since the seminal work of Chorin \cite{chorin1967numerical}. Volpe \cite{volpe1993performance} observed that the numerical error increases when the Mach-number is decreased, at a constant mesh and that the convergence rate deteriorates noticeably. Guillard and Viozat \cite{guillard1999behaviour} observe that an upwind space discretization leads to pressure fluctuations of the order of the Mach number $\eps$ while in the continuous case the pressure fluctuations are of order $\ec$. This difference originates from the upwinding terms, and more precisely from the eigenvalues of the Jacobian matrix whose order of magnitude is the sound velocity. The argument has been developed further in \cite{dellacherie2010analysis}.

Compressible codes also require an increasingly large Computational time as the incompressible regime gets closer. Indeed, the CFL stability condition for an explicit scheme reads $ \Delta t \leq \frac{\Delta x}{|\lambda^{\text{max}}|}$, where $\Dt$ is the time-step, $\Dx$ the space step, and $\lambda^{\text{max}}$ is the fastest characteristic wave and can be written $\lambda^{\text{max}}= u\pm c$, $u$ being the fluid velocity and $c$ the sound velocity. In scaled variables (see below for details on the scaling), the Mach number $\eps$ appears explicitly in the stability condition as follows: 
\begin{equation}
\Delta  \tilde  t \leq \frac{  \Delta \tilde x}{| \tilde \lambda^{\text{max}}|} = \frac{  \Delta \tilde x}{\max| \tilde u\pm\frac{ \tilde c}{\eps}|}=
							\eps\frac{ \Delta\tilde  x}{\max|\eps  \tilde u \pm  \tilde c|}, 
\label{eq:cflBM}
\end{equation} 
where the tildes denote scaled quantities and the sound speed is now written $\tilde c / \varepsilon$ where $\tilde c = {\mathcal O}(1)$. The time-step is therefore roughly proportional to the Mach number $\eps$ and is dramatically reduced when $\varepsilon$ is small.

The design of specific schemes for the small Mach-number regime has consequently triggered an abundant literature, following various tracks.
A first track consists in applying preconditioning methodologies. These methods have been initiated by the 'artificial compressibility' technique of Chorin \cite{choi1985application} and consist in multiplying the time-derivatives by a suitable matrix. They aim at modifying the eigenvalues of the compressible system in order to reduce the disparity between the acoustic and fluid wave speeds \cite{choi1985application, LiGu08, LiGu10, turkel1987preconditioned, van1991characteristic}. However, problems due to Computational instabilities related to the structure of the eigenvectors \cite{darmofal1996importance} and to the fact that the divergence-free constraint on the velocity is not always respected need to be dealt with. In most cases, these methods only apply to steady-state computations, since the time derivatives are modified. For non-stationary flows, dual time-stepping techniques can be introduced \cite{alkishriwi2006large} to recover time-accuracy. Working with the original compressible equations, \cite{gerbeau1997semi} construct a semi-implicit Roe-type solver by decomposing the Jacobian matrix into the fast and slow eigenvalues, the former being treated implicitly. In \cite{nerinckx2005mach}, the proposed scheme includes an implicit predictor convective step, followed by a semi-implicit corrector step.

A second track consist in focusing on the pressure equation. To this aim, a natural idea is to adapt classical incompressible schemes to the compressible case. The pressure-correction method SIMPLE \cite{karki1988pressure, patankar1980numerical} solves an elliptic pressure correction equation obtained via the mass conservation equation and the equation of state. In \cite{munz2003extension}, the elliptic pressure correction equation is obtained by introducing the pressure equation (derived from the energy equation) in the momentum equation. These methods respect the divergence constraint on the velocity but the formulation is not always conservative. In the ICE (Implicit Continuous Eulerian) method introduced by Harlow and Amsden and followers \cite{bonner2007, harlow1971numerical}, a splitting method is introduced between the explicit convective part and implicit acoustic part. However, the ICE method is not conservative and inaccurate shock speeds are observed. Klein \cite{klein1995semi} proposes a semi-implicit scheme which solves explicitly the leading order contribution of the pressure and the lower orders, implicitly. Other ways generating elliptic equations on the pressure can be found in \cite{kwatra2009method, MD01, NVD07, RVDM09, vidovic2006superlinearly, zienkiewicz1990compressible}.

A third track consists in using gauge (or Hodge) decomposition of the flow variables \cite{colella1999projection}. Indeed, the incompressible velocity between divergence free, it is tempting to decompose the compressible velocity into a divergence-free and a curl-free part. Semi-implicit time discretizations are used for the divergence-free part. The gauge decomposition was used in an earlier attempt to derive and AP-scheme \cite{degond2007gauge}. However, the method was too complex and never used. 

To some extent, our work belongs to the second class and relies on the introduction of a suitable elliptic equation on the pressure. However, it departs from previous work in that the problem is discretized in a single step, which reduces the Computational cost compared to predictor-corrector procedures, that the scheme is conservative and that the only equation solved implicitly is the elliptic equation, whose construction is extremely simple. 

More generally, AP-schemes have previously been proposed for neutron transport problems \cite{larsen1987asymptotic},  multiscale kinetic equations \cite{jin2001asymptotic}, hyperbolic heat equations \cite{gosse2002asymptotic}, relaxation limit of hyperbolic models \cite{lowrie2002methods}, plasmas in the quasi-neutral limit \cite{crispel2007asymptotic, degond2008analysis} or in the large magnetic field limit \cite{degond2009asymptotic}. 

The outline of  this paper is as follows. We first provide a semi-implicit AP time discretization of the compressible flow equations in section \ref{sec_time_semi}. Then, we derive the fully discrete (in time and space) AP-scheme at first order in section \ref{chapt:dis}. The construction of an elliptic equation on the pressure as well as the resolution of the scheme is detailed, and the outline of the extension to a second order scheme is given. Then, we perform the asymptotic analysis of the proposed scheme in section \ref{sec:asympt}, in order to show the AP property. Numerical results presented in section \ref{sec_num} provide a validation of the scheme in both the compressible and close-to-incompressible regimes. Finally, a conclusion is drawn at section \ref{sec_conclu}.


\setcounter{equation}{0}
\section{Time semi-discrete scheme}
\label{sec_time_semi}

We start with the Navier-Stokes equations (\ref{eq:NS1})-(\ref{eq:NS3}):
\begin{align}
  \partial_t \rho +  \nabla\cdot \rho \mathbf u &= 0,  \label{eq:NS1}\\
  \partial_t \rho \mathbf u +   \nabla\cdot (\rho \mathbf u\otimes \mathbf u) + \nabla p  &=  \nabla\cdot \left[ \rho \nu \left( ( \nabla \uv + \nabla \uv^T) - \frac{2}{3}(\nabla\cdot \uv) \mathbb{I} \right) \right] + \rho \mathbf f_{\text{ext}} , 
\label{eq:NS2}\\
  \partial_t \rho E +  \nabla\cdot (\rho H \mathbf u) &=   \nabla\cdot \left[ \frac{\lambda}{C_p} \nabla h \right]  + \rho \mathbf f_{\text{ext}}\cdot \mathbf u  , \label{eq:NS3} \\
W = \rho E &= \frac{1}{2}  \rho  u^2 +  \rho  h -  p, \label{eq:W0} 
\end{align}
 where $\rho$ is the density, $\uv$ is the velocity, $p$ is the pressure, $h$ is the enthalpy, $E$ is the total energy, $H = E+\frac{p}{\rho}$ is the total enthalpy,
 $\nu$ is the kinematic viscosity, $\nabla\uv^T$ is the transpose of the gradient of the velocity, $\mathbb{I}$ is the identity matrix, $\lambda$ is the conductivity, $C_p$ is the specific heat capacity, and $\mathbf f_{\text{ext}}$ represent external forces like gravity. The contribution of the term $\nabla\cdot \left[ \rho \nu\left( ( \nabla \uv + \nabla \uv^T) - \frac{2}{3}(\nabla\cdot \uv) \mathbb{I} \right) \right]\cdot \uv$ in the energy equation has been neglected, according to the models used in the {\small CEA} codes {\small FLICA4} \cite{flica4a, flica4b} and {\small CATHARE} \cite{morel97}, but could easily be added. We consider a general equation of state linking the density, the pressure and the enthalpy:
\begin{equation}
 \rho=\rho(p,h) .
\label{eq:eosfull}
\end{equation}

In this paper we deal with scaled equations. The scaling parameters $\rho_0$, $ p_0$, $u_0$, $x_0$ are introduced along with the scaled variables, denoted by a tilde.
\begin{eqnarray}
  \tilde \rho = \frac{\rho}{\rho_0}, \hspace{0.5cm} \tilde u =\frac{u}{u_0},  \hspace{0.5cm} \tilde p = \frac{p}{p_0}, \hspace{0.5cm} \tilde x = \frac{x}{x_0}, \hspace{0.5cm} \tilde E = \frac{\rho_0}{p_0}E, \hspace{0.5cm} \tilde h = \frac{\rho_0}{p_0}h.   
\end{eqnarray}
The scaled equation are the following (we will omit the tildes in the remainder of the paper):
\begin{align}
  \partial_t \rho +  \nabla\cdot \rho \mathbf u &= 0, \label{eq:NS1a}\\
  \partial_t \rho \mathbf u +   \nabla\cdot (\rho \mathbf u\otimes \mathbf u) + \frac{1}{\ec}\nabla p &= \frac{1}{Re} \nabla\cdot \left[ \rho  \left( ( \nabla \uv + \nabla \uv^T) - \frac{2}{3}(\nabla\cdot \uv) \mathbb{I} \right) \right] + \rho \mathbf f_{\text{ext}} ,
\label{eq:NS2a}\\
  \partial_t \rho E +  \nabla\cdot (\rho H \mathbf u) &=  \frac{1}{Re \cdot Pr} \bigtriangleup h + \ec \rho \mathbf f_{\text{ext}}\cdot \mathbf u ,  \label{eq:NS3a} \\
W = \rho E &= \frac{1}{2}\ec  \rho  u^2 +  \rho  h -  p, \label{eq:W} 
\end{align}
where the parameters resulting from the scaling are:
\begin{equation}
  \ec = \frac{\rho_0  u_0^2}{p_0}, \qquad  Re = \frac{u_0 x_0} {\nu},   \qquad Pr = \frac{\rho_0 \nu C_p}{\lambda}.
\label{eq:epslm}
\end{equation}
The parameter $\eps$ represents a global Mach number characterizing the flow and the nondimensionalisation. It is different from the local Mach number. The parameter $Re$ is the Reynolds number and $Pr$ is the Prandtl number.

For the sake of simplicity, the scheme is presented on the full Euler equations, which represent the convective part of the Navier-Stokes equations. The right-hand terms in the Navier-Stokes equations (\ref{eq:NS1})-(\ref{eq:NS3}) will be included later in explicit source terms, and the time semi-discretization will not be modified. 

The AP time semi-discrete scheme is written as follows:
\begin{align}
 & \frac{\rho^{n+1}-\rho^n}{\Delta t}  + \nabla\cdot \q^{n} = 0,\label{eq:enschfull1} \\
 & \frac{\q^{n+1}-\q^n}{\Delta t}  + \nabla\cdot (\frac{\q^n\otimes \q^n}{\rho^n} + \alpha p^n) +  \frac{1-\alpha\ec}{\ec} \bigtriangledown p^{n+1} = 0, \label{eq:enschfull2} \\
& \frac{W^{n+1}-W^n}{\Delta t} + \nabla\cdot H^n q^{n+1} = 0,  \label{eq:enschfull3} \\
& W^{n+1} = \rho^{n+1}e^{n+1} + \frac{1}{2}\ec  \rho^n  (u^n)^2 =  \rho^{n+1}  h^{n+1} -  p^{n+1} + \frac{1}{2}\ec  \rho^n  (u^n)^2, \label{eq:disW}
\end{align}
where $\Dt$ is the time-step, $t^n = n\Dt$ and the superscript '$n$' denotes the approximation of the variables at $t^n$, $\q=\rho\uv$, $W=\rho E$, and $\alpha$ is a small ad-hoc parameter independent of the Mach number and such that $\alpha \approx 1$, designed to prevent spurious oscillations in strong shock cases \cite{min}. The time discretization of the total energy $W=\rho E$ splits into an implicit evaluation of the internal energy $\rho e = \rho h -p$, and in an explicit evaluation of the kinetic energy. The discretization of the space derivatives is detailed in the next section (Section \ref{chapt:dis}).

\bigskip

Let us make a few comments on the proposed scheme. First, the scheme being conservative, we expect good shock properties in the compressible regime.
Then, we will see that the implicit treatment of the pressure in the momentum equation (\ref{eq:enschfull2}) is a key to the asymptotic preserving property (Section \ref{sec:approp}). 
An other noticeable feature is the implicit treatment of the momentum $\q$ in the energy equation, allowing us to construct an elliptic equation on the pressure. We now detail the resolution of the scheme and the construction of this elliptic equation.

\bigskip

The scheme can be solved through the following steps: 

 First, the density $\rho^{n+1}$ is obtained \textit{via} the resolution of the explicit continuity equation (\ref{eq:enschfull1}).

 An elliptic equation on the pressure is then solved. To construct this equation, the momentum equation (\ref{eq:enschfull2}) is rewritten as:
\begin{equation}
\q^{n+1} =\q^n - \Dt \bigtriangledown\cdot (\frac{\q^n\otimes \q^n}{\rho^n} + \alpha p^n) - \Dt \frac{1-\alpha\ec}{\ec} \bigtriangledown p^{n+1}.
\label{eq:mom2}
\end{equation}
This expression is inserted into the energy equation (\ref{eq:enschfull3}) and leads to:
\begin{equation}
W^{n+1} - \Dt^2\frac{1-\alpha\ec}{\ec} \bigtriangledown\cdot \left( H^n  \bigtriangledown p^{n+1} \right) = \phi(\rho^n, \q^n,W^n), \label{eq:ell1}
\end{equation}
where the right hand side $\phi$ is explicit and is equal to:
\begin{equation}
 \phi(\rho^n, \q^n,W^n) = W^n - \Dt \nabla\cdot H^n\q^n + \Dt^2 \nabla\cdot \left( H^n \bigtriangledown\cdot (\frac{\q^n\otimes \q^n}{\rho^n} + \alpha p^n) \right).
\end{equation}

Two cases can be considered : the specific case of a perfect gas equation of state, and the case of a general equation of state (EOS).
\paragraph{Perfect gas EOS}
For a perfect gas of polytropic constant $\gamma$, the internal energy reads $\rho e=\frac{1}{\gamma-1}p$. We can rewrite (\ref{eq:ell1}) as follows:

\begin{equation}
p^{n+1} - (\gamma-1)\Dt^2\frac{1-\alpha\ec}{\ec} \bigtriangledown\cdot \left( H^n  \bigtriangledown p^{n+1} \right) = \tilde\phi(\rho^n, \q^n,W^n), \label{eq:ellGP}
\end{equation}
with
\begin{equation}
  \tilde\phi(\rho^n, \q^n,W^n) = (\gamma-1)\phi(\rho^n, \q^n,W^n) - \frac{1}{2}(\gamma-1)\ec  \rho^n  (u^n)^2 .
\end{equation}

Equation (\ref{eq:ellGP}) is an elliptic equation on the pressure. It allows us to find the pressure $p^{n+1}$, and then $W^{n+1}$. 

\paragraph{General EOS}

For a general equation of state, the internal energy reads $\rho e = \rho h -p$. In this case, the following system has to be solved:
\begin{equation}
  \left\{
\begin{aligned}
&\rho^{n+1} h^{n+1} - p^{n+1} - \Dt^2\frac{1-\alpha\ec}{\ec} \bigtriangledown\cdot \left( H^n  \bigtriangledown p^{n+1} \right) = \tilde\phi'(\rho^n, \q^n,W^n)  \\ 
&\rho(  p^{n+1} ,h^{n+1} ) = \rho^{n+1} 
\end{aligned}
\right.,
\label{eq:ellGnal}
\end{equation}
where
\begin{equation}
 \tilde\phi'(\rho^n, \q^n,W^n)=W^n - \frac{1}{2}\ec  \rho^n  (u^n)^2 - \Dt \nabla\cdot H^n\q^n + \Dt^2 \nabla\cdot \left( H^n \bigtriangledown\cdot (\frac{\q^n\otimes \q^n}{\rho^n} + \alpha p^n) \right). 
\end{equation}
This still leads to an elliptic equation for the pressure, and the enthalpy is constrained by the value $\rho^{n+1}$ of the density found by the resolution of the explicit continuity equation. Solving this system allows us to find $p^{n+1}$, $h^{n+1}$ and $W^{n+1}$. 

 The momentum $\q^{n+1}$ is finally obtained \textit{via} the momentum equation (\ref{eq:mom2}), as $p^{n+1}$ is now known. Let us note that in (\ref{eq:mom2}) all terms are $O(1)$. Indeed, we have $ \frac{1-\alpha\ec}{\ec} \bigtriangledown p^{n+1} = O(1)$ due to the elliptic equation (\ref{eq:ellGnal}) which implies that $p^{n+1} =O(\ec) $ in the Sobolev space $H^2$ given the elliptic regularity theorem, and using appropriate boundary conditions. We thus get that $\nabla p^{n+1} =O(\ec) $ .

\medskip

 The proposed scheme presents two notable differences with the scheme for the isentropic equations presented in \cite{min}. First, the density is taken explicitly in the continuity equation. Then, the elliptic equation is obtained by the insertion of the momentum equation into the energy equation instead of into the continuity equation in the isentropic case. 
This difference is a consequence of the asymptotic analysis of the continuous full Euler equations (Section \ref{sec:asymptfulleuler}) where the divergence constraint on the velocity in the low Mach number regime is obtained from the energy equation.

\setcounter{equation}{0}
\section{Full time and space discretization}
\label{chapt:dis}

We present the full time and space discretization of the scheme for a first order scheme in a first part. Then we will extend the discretization to a second order scheme. We also insist on the centered space discretization of the implicit pressure.

\subsection{First order scheme}
\label{sec:firstorder}

In the finite volume framework, the first order space discretization of the scheme for a general multidimensional system on a structured or unstructured mesh is given by: 
\begin{align}
& \frac{\rho_i^{n+1}-\rho_i^n}{\Delta t}  + \sum \limits_{v \in \upsilon(i)}  \mathbf m_{iv}  \cdot
  \left[ \frac{\q_i^n + \q_v^n}{2} + D_{iv_\rho}^n \right] = S_{i_\rho}^n,  \label{eq:fd1} \\ 
 &\frac{\q_i^{n+1}-\q_i^n}{\Delta t}  + \sum \limits_{v \in \upsilon(i)}  \mathbf m_{iv} \cdot
  \left[ \beta_{iv}^n +   \frac{1-\alpha\ec}{2\ec} ( p_i^{n+1} +p_v^{n+1} ) \right] = \mathbf S_{i_q}^n, \label{eq:fd2}\\ 
&   \frac{\rho_i^{n+1} h_i^{n+1} - p_i^{n+1}  -W_i^n}{\Delta t} +\sum \limits_{v \in \upsilon(i)} \mathbf m_{iv}  \cdot
  \Bigg[ \frac{H_i^n \q_i^{n} + H_v^n \q_v^{n} }{2} + D_{iv_w}^n + \Dt \frac{H_i^n \mathbf S_{i_q}^n + H_v^n \mathbf S_{v_q}^n }{2} \nonumber \\
& \hspace{10em} -\Dt \frac{H_i^n}{2}  \sum \limits_{r \in \upsilon(i)} \mathbf m_{ir}  \cdot
  \left[ \beta_{ir}^n +   \frac{1-\alpha\ec}{2\ec} ( p_i^{n+1} +p_r^{n+1} ) \right] \label{eq:elliptic} \\
& \hspace{10em} -\Dt \frac{H_v^n}{2}  \sum \limits_{u \in \upsilon(v)} \mathbf m_{vu}   \cdot
  \left[  \beta_{vu}^n +   \frac{1-\alpha\ec}{2\ec} ( p_v^{n+1} +p_u^{n+1} ) \right] \Bigg]= S_{i_w}^n, \nonumber 
\end{align}
where, to simplify, we have introduced the notations:
\begin{align}
 \mathbf m_{iv} &= \frac{s_{iv}}{V_i}c_i^{iv}\n_{iv} , \\
 \beta_{iv}^n & = \frac{1}{2}(\frac{\q_i^n\otimes \q_i^n}{\rho_i^n} + \frac{\q_v^n\otimes \q_v^n}{\rho_v^n}) + \alpha \frac{p_i^n + p_v^n}{2} + \mathbf D_{iv_q}^n , 
\end{align}
\noindent  $\upsilon(i)$ is the set of neighbors of the cell $i$, $\n_{iv}$ is the unitary normal of the face between the $i$ and $v$ cells, $s_{iv}$ is the surface of this face, $V_i$ is the volume of the cell $i$, and $c_i^{iv}$ is $+1$ for an incoming normal of the face $iv$ into the cell $i$ and $-1$ for an outgoing normal, $ \mathbf D_{iv}^n =(D_{iv_\rho}^n,\mathbf D_{iv_q}^n,D_{iv_w}^n)$ is the upwinding between the $i$ and $v$ cells, taken at the time $n$, and detailed below.

Note that the energy equation (\ref{eq:enschfull3}) has been replaced by a discretization of the elliptic equation (\ref{eq:ellGnal}) on the pressure, the system so constituted being equivalent to the system (\ref{eq:enschfull1})-(\ref{eq:enschfull3}).

General source terms $S^n$ have been added and can include external forces such as gravity and the diffusive terms of the Navier-Stokes equations.

\paragraph{Upwinding} Centering the pressure term $\frac{1-\alpha\ec}{\ec} \nabla p^{n+1}$ in the spatial discretization is a crucial feature of the low Mach number scheme. It does not affect the stability as it is an implicit term. Then, the upwinding only concerns the explicit part of the flux in the equations (\ref{eq:fd1})-(\ref{eq:elliptic}) and the eigenvalues of the Jacobian matrix of the corresponding system are:
\begin{equation}
u_n-\sqrt{\alpha a_m^2} \quad,\quad |u_n| \quad,\quad u_n+\sqrt{\alpha a_m^2},
\end{equation}
where $u_n= \uv \cdot \n $ and  $a_m$ is the sound speed defined by:
\begin{equation}
 a_m=\left(\sqrt{\frac{\partial \rho}{\partial p}+\frac{1}{\rho}\frac{\partial \rho}{\partial h}}\right)^{-\ud}.
\label{eq:am}
\end{equation}
The CFL condition for the stability of the scheme is:
\begin{equation}
\Delta t \leq \frac{\Delta x}{max(u_n \pm\sqrt{\alpha a_m^2} )}.
\end{equation}
Therefore the time-step $\Dt$ does not depend on the Mach number $\eps$ contrary to a standard explicit method, as explained in the introduction. The time-step is based on the fluid velocity only : it does not take into account the acoustic velocity, which tends to infinity when the Mach number tends to zero and is responsible for the dramatic decrease of the time-step in the low Mach number regime. 
Also, the inaccuracy of explicit upwinding schemes is caused by the upwinding being based on the acoustic velocity, as recalled in the introduction and detailed in \cite{guillard1999behaviour}. To avoid introducing wrong pressure fluctuations, we must keep the parameter $\alpha$ small compared to $\frac{1}{\ec}$.

\medskip

In our scheme, a Lax-Friedrich upwinding is used. The term $\mathbf D_{iv}$ in the discretization (\ref{eq:fd1})-(\ref{eq:elliptic}) gives the upwinding between the cells $i$ and $v$ and its expression is:
\begin{equation}
\mathbf D_{iv}\cdot \n_{iv}= 
  \begin{pmatrix}
D_{iv_\rho} \\
   \mathbf D_{iv_q} \\
    D_{iv_w}
  \end{pmatrix} \cdot \n_{iv}
=-  \ud(\lambda_n^{\text{max}})_{iv}(\V_v-\V_i),
\end{equation}
where $\V=(\rho, \q, W)$ is the vector of conservative variables and 
\begin{equation}
 (\lambda_n^{\text{max}})_{iv} = \max \Big( \quad |u_n|_i+\sqrt{\alpha a_m^2}_i \quad,\quad |u_n|_v + \sqrt{\alpha a_m^2}_v \quad    \Big).
\label{eq:lmax}
\end{equation}

\paragraph{Resolution of the discrete system}
\label{sec:resol1erordre}

Let us detail the steps in the resolution of the scheme.

First, the mass equation (\ref{eq:fd1}) can be solved explicitly, and $\rho^{n+1}$ is obtained.

Then we solve the elliptic equation (\ref{eq:elliptic}). For a perfect gas EOS, this elliptic equation is a linear system on the pressure and can be solved by inverting the system. In the case of a general EOS, the system constituted by the elliptic equation (\ref{eq:elliptic}) and the equation of state is solved by means of a Newton method where the unknowns are the pressure and the enthalpy. 

We will note $(p^{(q)}$, $h^{(q)})$ the pressure and enthalpy found by the $q^{th}$ iteration in the Newton method in order to find $(p^{n+1}$, $h^{n+1})$ at the time $t^{n+1}$. Two iterations $q$ and $q+1$ of the Newton method are linked by the following relation:
\begin{equation}
 \begin{pmatrix}
 p^{(q+1)}\\
h^{(q+1)}
\end{pmatrix} 
 = \begin{pmatrix}
 p^{(q)}\\
h^{(q)}
\end{pmatrix}  - f'(p^{(q)},h^{(q)})^{-1}  f(p^{(q)},h^{(q)}),
\label{eq:newton}
\end{equation}
where the algorithm is initialized with $(p^{(0)},h^{(0)})=(p^n, h^n)$. The function $ f$ is a vector defined as: 
\begin{equation}
 f(p^{(q)},h^{(q)}) = (\ec f_1(p^{(q)},h^{(q)}), f_2(p^{(q)},h^{(q)})),
\end{equation}
The first component $f_1$ comes from the elliptic equation (\ref{eq:elliptic}) and the second component $f_2$ expresses the condition over the pressure and the enthalpy given by $\rho^{n+1}$:
\begin{equation}
f_2(p^{(q)},h^{(q)}) =  \rho^{n+1} - \rho(p^{(q)},h^{(q)}).
\end{equation}
In practice, the first component of $f$ is $\ec f_1$ in order to avoid the division by the small parameter $\ec$ in the term $\frac{1-\alpha\ec}{2\ec}$.

The matrix $f'(p^q,h^q)$ in (\ref{eq:newton}) is the following:

\begin{equation}
  f'(p^q,h^q) = 
\begin{pmatrix}
 \ec \frac{\partial f_1}{\partial p} & \ec \frac{\partial f_1}{\partial h} \\
 \frac{\partial \rho}{\partial p} & \frac{\partial \rho}{\partial h}
\end{pmatrix}
.
\end{equation}

Solving the elliptic equation allows us to find $p^{n+1}$, $h^{n+1}$ and $W^{n+1}$. 
Finally, the momentum equation (\ref{eq:fd2}) is solved to obtain $\q^{n+1}$ and $\uv^{n+1}$.

\subsection{Second order scheme}

The first order scheme being too diffusive,we propose a second-order space discretization of the scheme.

\medskip

In the first order system, the full time and space discretization (\ref{eq:fd1})-(\ref{eq:elliptic}) could be written as:
\begin{equation}
 \frac{\V_i^{n+1}-\V_i^n}{\Dt} + \sum \limits_{v \in \upsilon(i)} \Phi(\V_i^{n,n+1}, \V^{n,n+1}_v) =0 ,
\end{equation}
where $\Phi$ is the numerical flux and $\V_i$ is the vector of conservative variables in the center of the cell $i$. The second order space discretization consists in evaluating the numerical flux $\Phi$ in the reconstructed and limited states $\Vt_{iv}^L$ and $\Vt_{iv}^R$, which correspond to the vectors of conservative variables $\V_i$ and $\V_v$ on the face between the cells $i$ and $v$. We thus replace $\Phi(\V_i^{n}, \V^{n}_v)$ by $\Phi \left((\Vt_{iv}^L)^{n}, (\Vt_{iv}^R)^{n}\right)$. The minmod limiter is used to avoid spurious oscillations. 

\smallskip

On a two-dimensional Cartesian mesh, the reconstructed and limited states $\Vt_{i+\ud,j}^L$ and $\Vt_{i+\ud,j}^R$ are given by the following expressions (\cite{lin1993upwind, GodlewskiRaviart}):
\begin{align}
  \Vt_{i+\ud,j}^L &= \V_{i,j} + \ud \minmod( \V_{i,j}-\V_{i-1,j}, \V_{i+1,j} - \V_{i,j}) \, , \label{eq:recons1}\\
  \Vt_{i+\ud,j}^R &= \V_{i+1,j} - \ud \minmod( \V_{i+2,j}-\V_{i+1,j}, \V_{i+1,j} - \V_i,j)\, , \label{eq:recons2}
\end{align}
where the minmod function is:
\begin{equation}
 \minmod(x,y) = \ud [\text{sign}(x) +\text{sign}(y)] min(|x|, |y|).
\end{equation}

The upwinding of the scheme is now given by:
\begin{equation}
\mathbf D_{iv}\cdot \n_{iv}= 
  \begin{pmatrix}
D_{iv_\rho} \\
   \mathbf D_{iv_q} \\
    D_{iv_w}
  \end{pmatrix} \cdot \n_{iv}
=-  \ud(\lambda_n^{\text{max}})_{iv}\left(\Vt_{iv}^R -\Vt_{iv}^L \right),
\label{eq:upwind2nd}
\end{equation}
where $(\lambda_n^{\text{max}})_{iv}$ is still given by equation (\ref{eq:lmax}).

\medskip

To solve this scheme, the reconstructed and limited states $(\Vt_{iv}^L)^n$ and $(\Vt_{iv}^R)^n$ are first calculated from the conservative vector $\V^n$. In addition, we also obtain the corresponding pressure. These values are used to evaluate the numerical fluxes, the upwinding and the source terms. The mass equation is first solved explicitly and $\rho^{n+1}$ is found. Then, the elliptic equation is solved by means of a Newton method, as explained for the first-order scheme. The momentum equation is then solved and $\V^{n+1}$ is obtained.

\medskip

A second-order discretization in time was intended \textit{via} a Runge-Kutta method combined with Crank-Nicolson (RK2CN). However the scheme tended to generate spurious oscillations if the pressure was fully implicit ($\alpha=0$). To date, the problem has not been solved by the authors. As the full implicit treatment of the pressure is necessary in the low Mach regime, the time and space second-order scheme could not be used. In the next part of the article, the ``second-order scheme'' will thus refer to the first-order in time and second-order in space scheme.

\setcounter{equation}{0}
\section{Asymptotic preserving property}
\label{sec:asympt}

Let us now show that the proposed scheme is asymptotic preserving. The asymptotic preserving property has been defined in the introduction.
We first recall the asymptotic study of the full Euler equations as the methodology is used in the study of the asymptotic preserving property of the scheme.

\subsection{Asymptotic analysis of the continuous Euler equations}
\label{sec:asymptfulleuler}
 
Let us now investigate the limit of the full Euler equations when $\eps\to 0$. 
The method differs from the isentropic case \cite{min} as the condition on the divergence of the velocity in the low Mach number regime is obtained via the energy equation instead of the continuity equation. This is a consequence of the density depending both on pressure and enthalpy (\ref{eq:eosfull}).
The analysis below extends the asymptotic analysis led by Klein in \cite{klein1995semi, klein2001asymptotic} for the full Euler equations to a general equation of state.

\medskip

If we write the expansions of the variables $\rho$, $p$, $\uv$, $H$ and $W$ in powers of the Mach number $\eps$, e.g. $ \rho = \rho_0 +  \eps \rho_{(1)}+ \ec \rho_{(2)}+\ldots$, and insert them in the full Euler equations, the leading order equations are: \vspace{-1em}
\begin{align}
 & \partial_t \rho_{(0)} + \nabla\cdot \left( \rho_{(0)} \uv_{(0)} \right) = 0, \label{eq:massleading}\\
 & \nabla  p_{(0)} = 0, \label{eq:gradp}\\
 & \partial_t W_{(0)} + \nabla\cdot\left( \rho_{(0)} H_{(0)} \uv_{(0)} \right) = 0, \label{eq:wleading}
\end{align}
and the second order equation for the momentum is:
\begin{equation}
 \partial_t (\rho\uv)_{(0)} + \nabla(\rho_{(0)}\uv_{(0)}\otimes\uv_{(0)}) + \nabla p_{(2)}=0. 
\label{eq:momordre2}
\end{equation}
The variable $ p_{(2)}$ is a dynamic pressure as it is directly linked to the speed of the fluid, while $p_{(0)}$ is a thermodynamic pressure as it appears in the energy equation. Eq.(\ref{eq:gradp}) yields that $p_{(0)}$ is independent of space. We assume that the boundary conditions are chosen such that the constant $p_{(0)}$ is independent of time.
 As the parameter $\ec$ appears in the expression (\ref{eq:W}) of $W$, at leading order we have:
 \begin{equation}
W_{(0)} = \rho_{(0)} e_{(0)} \quad \text{and} \quad \rho_{(0)} H_{(0)} = \rho_{(0)} e_{(0)}+p_{(0)}, \quad \text{\textit{ie}} \quad H_{(0)} = h_{(0)}.
\end{equation}

We first recall the simpler case of a perfect gas, then extend the analysis to a general equation of state.

\paragraph{Perfect gas case} 

For a perfect gas with a constant $\gamma$, we have $\rho e = \frac{1}{\gamma -1} p$. Therefore $W_{(0)} = \frac{1}{\gamma -1} p_{(0)} $  and $\rho_{(0)} H_{(0)} =\frac{\gamma}{\gamma -1} p_{(0)} $ are independent of space due to (\ref{eq:gradp}), and independent of time. The leading order of the energy equation (\ref{eq:wleading}) gives the divergence condition on the velocity in the zero Mach number limit:
\begin{equation}
\nabla \cdot \uv_{(0)} = 0.
\end{equation}

\paragraph{General EOS case}

We drop the subscript $(0)$ for simplicity. With $W=\rho h-p$, $H=h$ and (\ref{eq:gradp}), we get:
\begin{equation}
 \partial_t W + \nabla \cdot (\rho h \uv) = (\partial_t  + \uv\cdot \nabla )(\rho h) + \rho h ( \nabla \cdot \uv) =0.
\end{equation}
Now, $\rho h = \rho h(\rho, p)$ for a general EOS, and we get 
\begin{align}
(\partial_t  + \uv\cdot \nabla )(\rho h) &= \frac{\partial \rho h}{\partial \rho} (\partial_t  + \uv\cdot \nabla )\rho + \frac{\partial \rho h}{\partial p} (\partial_t  + \uv\cdot \nabla )p \\
 &= -\rho \frac{\partial \rho h}{\partial \rho} \nabla \cdot \uv,
\end{align}
thanks to (\ref{eq:gradp}) and the assumption that $\partial_t p=0$.
We collect the above equations and get
\begin{equation}
 (\rho h - \rho \frac{\partial \rho h}{\partial \rho}) \nabla \cdot \uv =0,
\end{equation}
or, since $\rho h - \rho \frac{\partial \rho h}{\partial \rho} = -\rho^2  \frac{\partial  h}{\partial \rho}$,
\begin{equation}
 \frac{\partial  h}{\partial \rho} \nabla \cdot \uv =0.
\end{equation}
With the assumption that $ \frac{\partial  h}{\partial \rho} \neq 0$, we get the incompressibility condition:
\begin{equation}
\nabla \cdot \uv = 0.
\label{eq:divusansp}
\end{equation}
The divergence of the velocity being zero, the mass equation (\ref{eq:massleading}) becomes
\begin{equation}
 \partial_t \rho +  \uv \cdot \bigtriangledown \rho = 0,
\label{eq:eqsurrho}
\end{equation}
which expresses that the density is constant along a trajectory of any fluid element. By contrast, in the isentropic case, the low Mach number limit leads to a constant density in space.

\medskip

Equations (\ref{eq:gradp}), (\ref{eq:momordre2}), (\ref{eq:divusansp}) and (\ref{eq:eqsurrho}) form the incompressible limit of the Euler equations. 
Klainerman and Majda in \cite{klainerman1981singular, klainerman1982compressible}, then Metivier and Schochet in \cite{metivier2001incompressible} have shown that the solution of the compressible Euler equations converges towards the solution of the incompressible Euler equations when $\eps$ tends to zero.

\subsection{Study of the asymptotic preserving property of the scheme}
\label{sec:approp}

Let us now show that the proposed scheme is asymptotic preserving. We expose the reasoning on the time semi-discrete scheme (\ref{eq:enschfull1})-(\ref{eq:enschfull3}) for the sake of simplicity and readability. The extension to the full time and space discretization is straightforward. 

 To show the asymptotic preserving property, we have to write the limit discrete scheme ${\cal{M}}^0_\Delta$ when $\eps\to 0$ and show that it is consistent with the continuous limit model ${\cal{M}}^{0}$ at $\eps= 0$.

The continuous limit model ${\cal{M}}^{0}$ is the following:
\begin{equation}
  \left\{
\begin{aligned}
 & \partial_t \rho_{(0)} +  \nabla \cdot ( \rho_{(0)} \uv_{(0)}) = 0,\\
 & \nabla  p_{(0)} = 0, \\
& \partial_t (\rho\uv)_{(0)} + \nabla\cdot(\rho_{(0)}\uv_{(0)}\otimes\uv_{(0)}) + \nabla \pi=0, \\
& \partial_t W_{(0)} + \nabla  \cdot ( \rho_{(0)} H_{(0)}\uv_{(0)}) =0, \\
& H_{(0)} = h_{(0)}, \quad W_{(0)} = \rho_{(0)} e_{(0)} = \rho_{(0)} h_{(0)} - p_{(0)},
\end{aligned}
\right.
\label{eq:continouslimitsystem}
\end{equation}
where $\pi$ is a dynamic pressure and $p_{(0)}$ a thermodynamic pressure, and under the assumption that $p_{(0)}$ is independent of time.

We introduce the expansions in powers of $\eps$ in the semi-discrete scheme (\ref{eq:enschfull1})-(\ref{eq:enschfull3}) in the same way as in the asymptotic analysis of the continuous case (Section \ref{sec:asymptfulleuler}). Considering the leading order equations and the equation of order two for the momentum equation, we obtain the discrete limit system ${\cal{M}}^0_\Delta$:
\begin{equation}
  \left\{
\begin{aligned}
 & \frac{\rho_{(0)}^{n+1}-\rho_{(0)}^n}{\Delta t}   +   \nabla \cdot( \rho_{(0)}^{n} \uv_{(0)}^{n})  = 0,\\
 & \nabla p_{(0)}^{n+1}=0, \\
& \frac{(\rho\uv)_{(0)}^{n+1}-(\rho\uv)_{(0)}^n}{\Delta t}  + \nabla\cdot (\rho_{(0)}^n\uv_{(0)}^{n}\otimes\uv_{(0)}^{n})  +  \nabla p_{(2)}^{n+1} = 0,\\
&  \frac{(\rho e)_{(0)}^{n+1}-(\rho e)_{(0)}^n}{\Delta t} + \nabla\cdot( h_{(0)}^n \rho_{(0)}^{n+1}\uv_{(0)}^{n+1} ) = 0,\\
& W_{(0)}^{n+1} = (\rho_{(0)} e_{(0)})^{n+1} =  (\rho_{(0)} h_{(0)})^{n+1} - p_{(0)}^{n+1}.
\end{aligned}
\right.
\label{eq:sdislimitsystem1}
\end{equation}

System (\ref{eq:sdislimitsystem1}) is clearly consistent with system (\ref{eq:continouslimitsystem}). Therefore, the scheme is asymptotic preserving. 

Nonetheless, we show directly that (\ref{eq:sdislimitsystem1}) is also consistent with the incompressibility constraint, namely that

\paragraph{\underline{Proposition:}} $\nabla \cdot \uv_{(0)}^{n+1} = O(\Dt)\,, $ where $O(\Dt)$ is independent of $\eps$.

\bigskip

\underline{\textit{Remark 1:}} From now on, we drop the subscript $(0)$ and $O(\Dt)$ will denote terms estimated by $C \Dt$ with $C$ independent of $\eps$.

\underline{\textit{Remark 2:}} From  (\ref{eq:sdislimitsystem1}), we deduce in particular that 
\begin{align}
 & \rho^{n+1}=\rho^n  + O(\Dt),\label{eq:ap5}\\
& \uv^{n+1}=\uv^n + O(\Dt),\label{eq:ap7}\\
&  (\rho h)^{n+1} =(\rho h)^n + p^{n+1} - p^n + O(\Dt),\label{eq:ap8}
\end{align}
with $O(\Dt)$ independent of $\eps$. From (\ref{eq:ap7}), we deduce that
\begin{equation}
\nabla \cdot  \uv^{n+1}= \nabla \cdot \uv^n + O(\Dt).\label{eq:ap9}
\end{equation}
However, even if $\nabla \cdot \uv^0 =0$, this does not prove that $\nabla \cdot \uv^{n+1}= O(\Dt)$, since summing over all time steps will lead to  $\nabla \cdot \uv^{n+1}= O(1)$. 
Therefore, we need to show directly that $\nabla \cdot \uv^{n+1}= O(\Dt)$. The proof is similar as in the continuous case.

\underline{\textit{Remark 3:}} From the second equation of (\ref{eq:sdislimitsystem1}), we deduce that $p^{n+1}$ is independent of $x$. We assume that the boundary conditions are such that $p^{n+1}$ is also independent of $n$, \textit{i.e.} $p^{n+1} = p^n=...=p^1=p^0$.

\bigskip

\textbf{\underline{Proof:}} 

 We write the fourth equation of (\ref{eq:sdislimitsystem1}) as
\begin{equation}
  \frac{(\rho h)^{n+1}-(\rho h)^n}{\Delta t}  - \frac{p^{n+1}-p^n}{\Delta t} + \uv^{n+1} \cdot \nabla( h^n \rho^{n+1} )+ h^n \rho^{n+1} \nabla\cdot \uv^{n+1}  = 0.
\label{eq:proof1}
\end{equation}
Since 
\begin{equation}
 p^{n+1} - p^n=0, \quad (\rho h)^{n+1} = (\rho h) (\rho^{n+1}, p^{n+1}), \quad (\rho h)^n = (\rho h)(\rho^n,p^n),
\end{equation}
we have, using (\ref{eq:ap7}) and the first equation of (\ref{eq:sdislimitsystem1}):
\begin{align}
  \frac{(\rho h)^{n+1}-(\rho h)^n}{\Delta t}  - \frac{p^{n+1}-p^n}{\Delta t}  
& = \frac{1}{\Dt} \Bigg( \frac{\partial (\rho h)}{\partial \rho}(\rho^n, p^n)(\rho^{n+1} -\rho^n) +  \frac{\partial (\rho h)}{\partial p}(\rho^n, p^n)(p^{n+1} -p^n)  \nonumber\\
 & \hspace{5em} + O\left((\rho^{n+1} -\rho^n)^2  \right) + O\left((p^{n+1} -p^n)^2  \right) \Bigg) \nonumber  \\
& = \frac{\partial (\rho h)}{\partial \rho}(\rho^n, p^n) \frac{\rho^{n+1}-\rho^n}{\Delta t}  + O(\Dt) \nonumber \\
& = - \frac{\partial (\rho h)}{\partial \rho}(\rho^n, p^n) \left[ \uv^n \cdot \nabla  \rho^n + \rho^n \nabla \cdot \uv^n  \right] + O(\Dt) \nonumber \\
& = - \frac{\partial (\rho h)}{\partial \rho}(\rho^n, p^n) \left[ \uv^{n+1} \cdot \nabla  \rho^n + \rho^n \nabla \cdot \uv^{n+1}  \right] + O(\Dt).
\label{eq:proof2} 
\end{align}
Similarly, using (\ref{eq:ap5}) we have
\begin{align}
 \uv^{n+1}\cdot \nabla (\rho^{n+1}h^n) &= \uv^{n+1}\cdot \nabla (\rho^{n}h^n)  + O(\Dt) \nonumber  \\
 &= \uv^{n+1} \cdot \Bigg[ \frac{\partial (\rho h)}{\partial \rho}(\rho^n, p^n) \nabla \rho^n +  \frac{\partial (\rho h)}{\partial p}(\rho^n, p^n) \nabla p^n \Bigg] + O(\Dt) \nonumber \\
&=  \frac{\partial (\rho h)}{\partial \rho}(\rho^n, p^n)\uv^{n+1} \cdot\nabla \rho^n+ O(\Dt) .
\label{eq:proof3}
\end{align}
Adding (\ref{eq:proof2}) and (\ref{eq:proof3}) in view of (\ref{eq:proof1}) leads to 
\begin{equation}
 \left( \rho \left[ h - \frac{\partial (\rho h)}{\partial \rho} \right] \right)^n \nabla \cdot \uv^{n+1} = O(\Dt),
\end{equation}
or 
\begin{equation}
\left( \frac{\partial h}{\partial \rho}\right)^n \nabla \cdot \uv^{n+1} = O(\Dt).
\end{equation}
With $\left( \frac{\partial h}{\partial \rho}\right)^n \neq 0$, we deduce that $ \nabla \cdot \uv^{n+1}= O(\Dt)$ which ends the proof.

The proof of the asymptotic preserving property for the fully discrete scheme follows the same methodology and is left to the reader.

\setcounter{equation}{0}
\section{Numerical results}
\label{sec_num}

In this part we provide numerical results for the second-order asymptotic preserving scheme, the first-order scheme being too diffusive. We first test the accuracy and the convergence order of the scheme on the colliding acoustic pulses test-case, then study the behavior of the scheme in the compressible regime with shock tubes test-cases, using the Euler equations. At last, we test the behavior of the scheme at low Mach number with the well-known test-cases of the backward facing step and the lid driven cavity, modeled by the full Navier-Stokes equations, and the test-case of the heat-driven cavity, which uses the energy equation. The results are compared to the results of the Low Mach Roe scheme described in \cite{dellacherie2010analysis, dellacherie2010influence}, using the OVAP code to run the simulations \cite{ovap}. The Low Mach Roe scheme is an incompressible solver and has been the object of previous validation.

\subsection{Colliding acoustic pulses}

This test-case proposed in \cite{klein1995semi} consists in two acoustic pulses, a right-running pulse and a left-running pulse. The pulses first collide and superpose, with a maximum of pressure at $t=0.815$s, then separate to return to their initial configuration at $T=1.63$s. The boundary conditions are periodic. The Mach number is $\eps = \frac{1}{11}$ and we use a perfect gas of constant $\gamma=1.4$, the equation of state being $\rho = \frac{\gamma}{\gamma-1}\frac{p}{h}$. A one-dimensional domain $[-L,L]$ is considered, with $L = \frac{2}{\eps}$, and is discretized into 220 cells. The time-step is $\Dt=1 \times 10^{-3}$s. The parameter $\alpha$ in the numerical scheme is taken as $\alpha=0$. The initial data for this case are:
\begin{eqnarray}
\rho(x,0) = \rho_0 + \ud \eps \rho_1 (1-cos(2\pi \frac{x}{L})),  & \rho_0=0.995, &  \rho_1 = 2.0\\
p(x,0) =p_0 + \ud \eps p_1 (1-cos(2\pi \frac{x}{L})),   &  p_0=1.0, & p_1=2\gamma ,\\
u(x,0) = \ud \text{sign}(x) u_0 (1-cos(2\pi \frac{x}{L})),   & u_0 = 2 \sqrt{\gamma}. &  
\end{eqnarray}

\begin{figure}[h!]
\begin{center}
\subfigure[t=0.185s]{\includegraphics[width=0.48\textwidth,trim=30 50 80 80,clip]{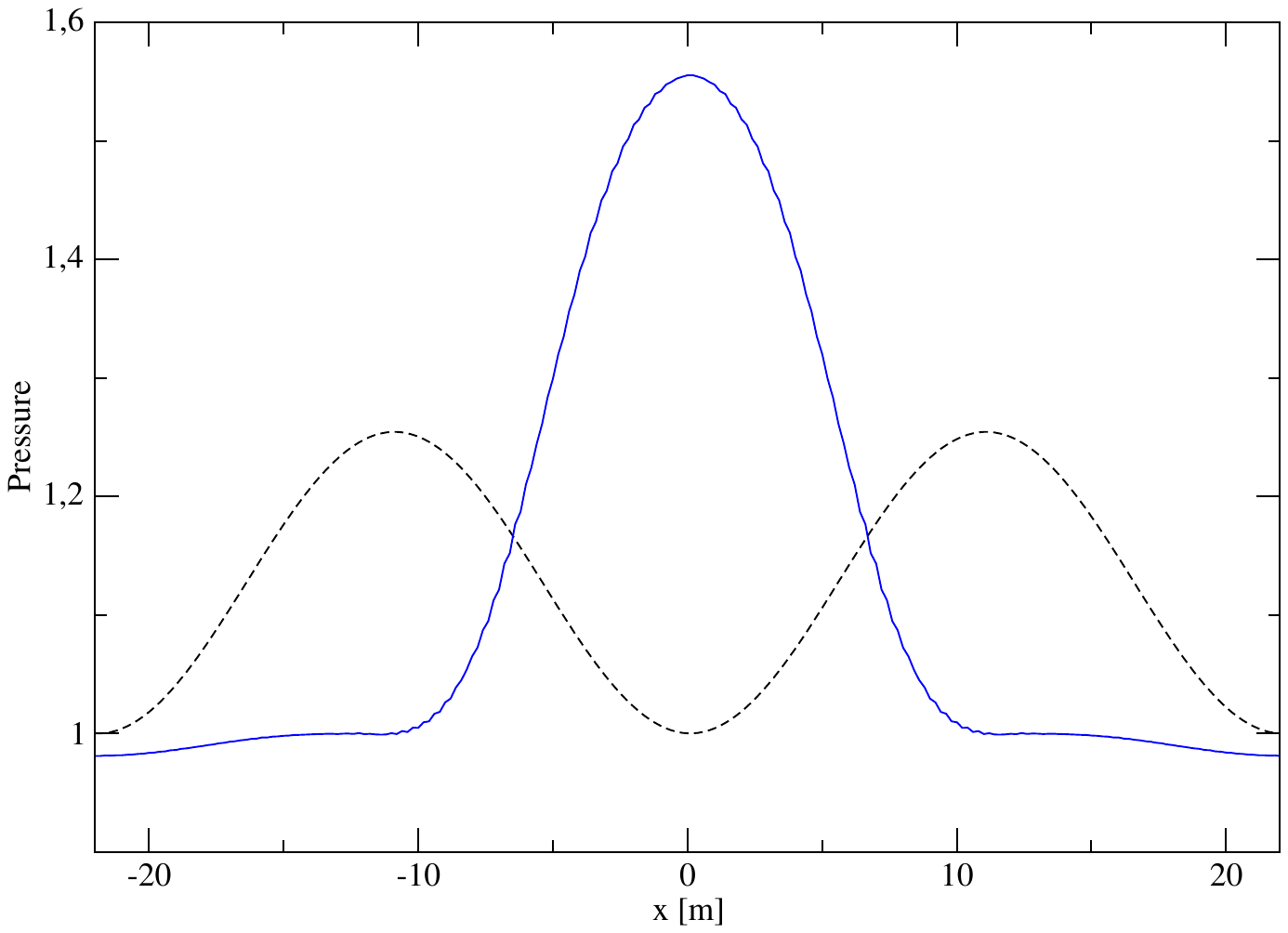}}
\subfigure[t=1.63s]{\includegraphics[width=0.48\textwidth,trim=30 50 80 80,clip]{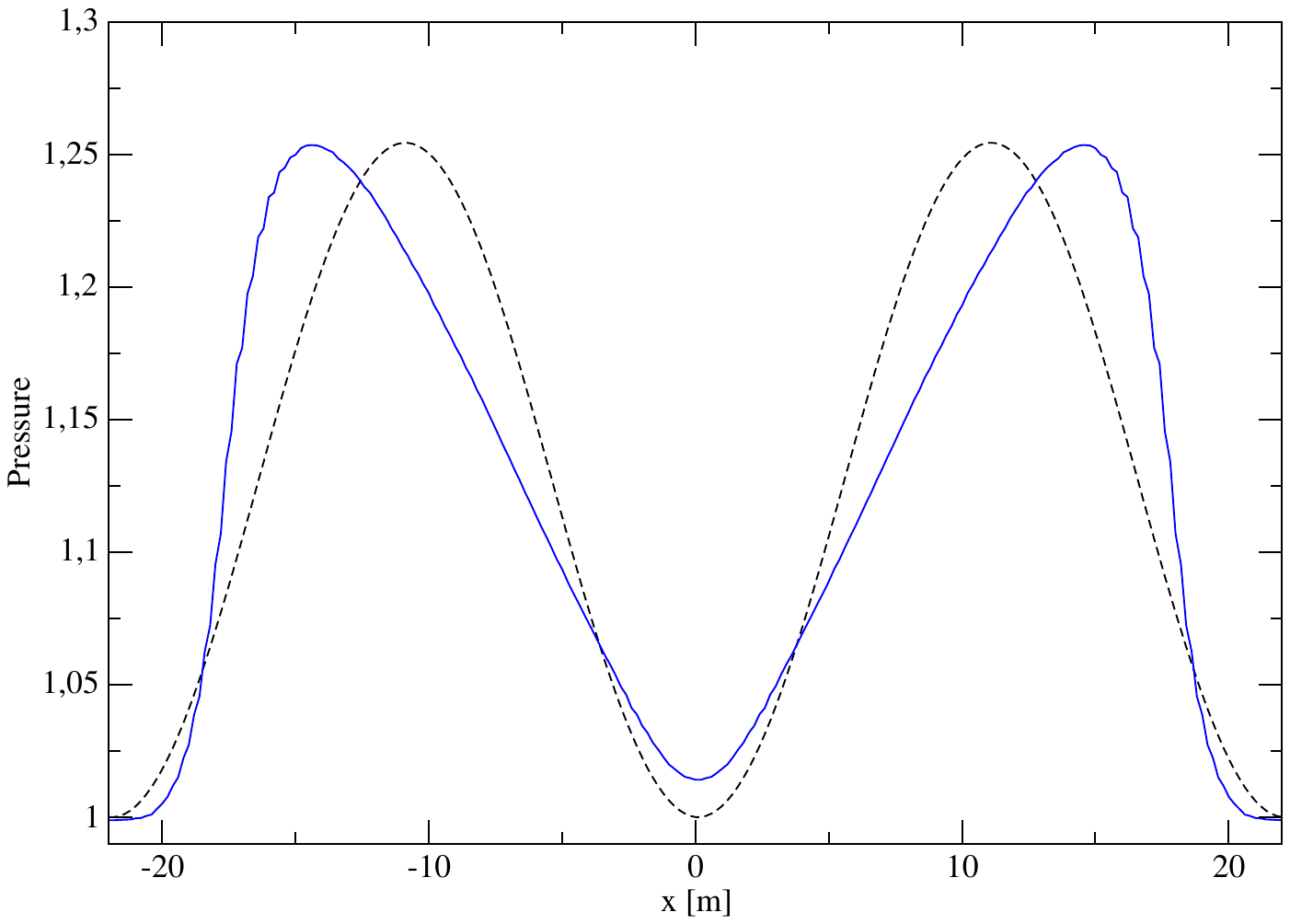}}
\caption{Pressure profile for the colliding pulses test-case. The initial profile is in dashed line, and the solid line gives the result of the second-order scheme for two different times.}
\label{fig:colliding}
\end{center}
\end{figure} 

The pressure profile computed by the second-order scheme is compared to the initial condition on fig. \ref{fig:colliding}. At $t=0.815$s, the pressure reaches a maximum value as the two pulses are superposed. At $t=1.63s$, the pulses are separated from each other again. As explained in \cite{klein1995semi}, weakly nonlinear acoustic effects distort the final profile as shocks are beginning to form in the vicinities of the locations $x=\pm 18.5$. 

\paragraph{Convergence tests}

We check that the scheme has indeed a second order convergence in space. We calculate the error between the solution $p$ obtained for the pressure with $N=100$, 200 and 400 cells with a reference solution $p_{ref}$ calculated with $N_{ref}=3200$ cells. In order to check the spatial convergence only, the time-step is taken as $\Dt = 0.05\times \Dx^2$.

The error $||E||_{L^1}$ is the discrete $L^1$ norm of the difference between the solution $p$ and the reference solution $p_{ref}$:
\begin{equation}
 ||E||_{L^1} = \frac{\sum_{j=1}^{N_{ref}} |p(x_j)-p_{ref}(x_j)|}{\sum_{j=1}^{N_{ref}} |p_{ref}(x_j)|},
\label{eq:cvtests}
\end{equation}
where $p(x_j)$ is calculated by linear interpolation when $x_j$ is not a discretization point for the discrete solution, as $p$ has been computed with less cells than $p_{ref}$.

\begin{figure}
\begin{center}
\includegraphics[width=0.68\textwidth,trim=30 50 80 80,clip]{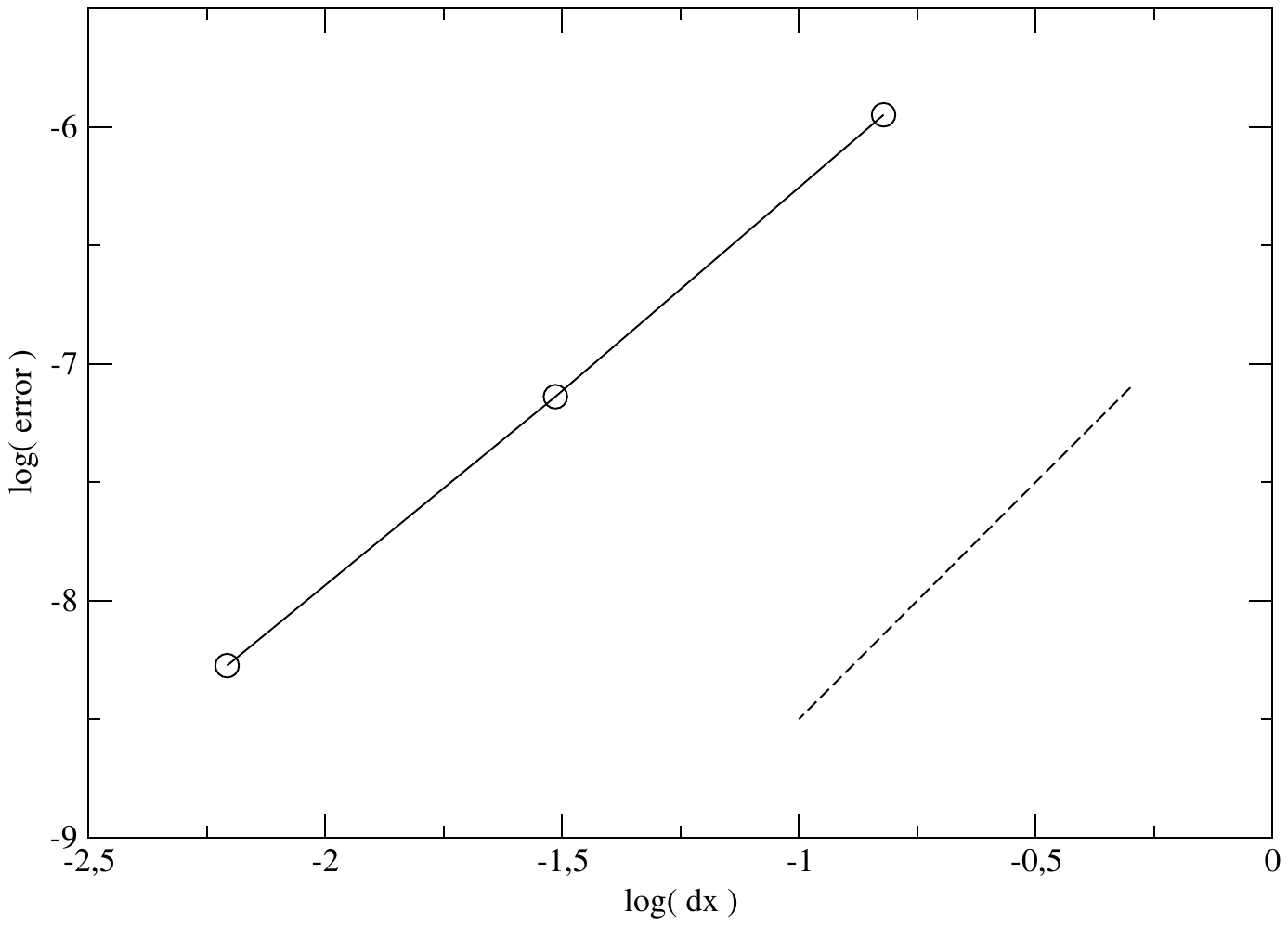}
\caption{Convergence test for the first order in time and second order in space scheme on the colliding pulses test-case. The solid line is the log of the $L^1$ error as a function of the log of $\Dx$. The dashed line is a line of slope 2.}
\label{fig:testCVFstordertime}
\end{center}
\end{figure}

The results have been computed with $\alpha=10$ to avoid spurious oscillations due to the shocks forming in the vicinities of the locations $x=\pm 18.5$.
 The $L^1$ norm of the relative error between the reference solution and the results for 100, 200 and 400 cells is given in table \ref{table:cvColliding}. The logarithm of the error as a function of the logarithm of the space step $\Dx$ is plotted on fig. \ref{fig:testCVFstordertime}. We indeed have a second order convergence in space.
\begin{table}[ht!]
\begin{center}
  \begin{tabular}{|c|c|c|c|}
\hline
Cells & $\Dx$ & $\Dt= 0.05\times \Dx^2$ & $||E||_{L^1}$ \\\hline
100 & 0.44 & $9.68 \times 10^{-3}$ & $2.61\times 10^{-3}$ \\
200 & 0.22 & $2.42 \times 10^{-3}$ & $7.94\times 10^{-4}$ \\
400 & 0.11 & $6.05 \times 10^{-4}$ & $2.55\times 10^{-4}$ \\ \hline
\end{tabular}
\caption{Colliding pulses test-case. $L^1$ norm of the relative error between the reference solution computed with 3200 cells and the numerical results for 100, 200 and 400 cells.}
\label{table:cvColliding}
\end{center}
\end{table}

\subsection{Shock tube problems}

\paragraph{Sod shock tube}

This shock tube test-case has been proposed in \cite{sod1978survey}. The initial state is divided in a left part $0\leq x \leq 0.5$ and a right part $0.5 < x \leq 1$, the initial values being given in table \ref{table:tacSod}.
\begin{table}[b]
\begin{center}
 \begin{tabular}{|c|c|c|c|}
\hline
      & p  & h  & u  \\ \hline
Left &  1   & 3.5 & 0      \\ \hline
Right &  0.1  &  2.8    &  0 \\ \hline
\end{tabular}
\caption{Sod shock tube. Initial conditions for the pressure, the enthalpy and the velocity.}
\label{table:tacSod}
\end{center}
\end{table} \\
We use a perfect gas of equation of state $\rho = \frac{\gamma}{\gamma-1}\frac{p}{h}$, with a constant $\gamma =1.4$, Neumann boundary conditions, and a mesh with 100 cells, with $\Delta x = 0.01$ and $\Dt = 0.001$. 
The Mach number is $\eps = 1$ and the parameter $\alpha$ is zero. 

\begin{figure}
\begin{center}
\subfigure[Density]{\includegraphics[width=0.48\textwidth,trim=30 50 80 80,clip]{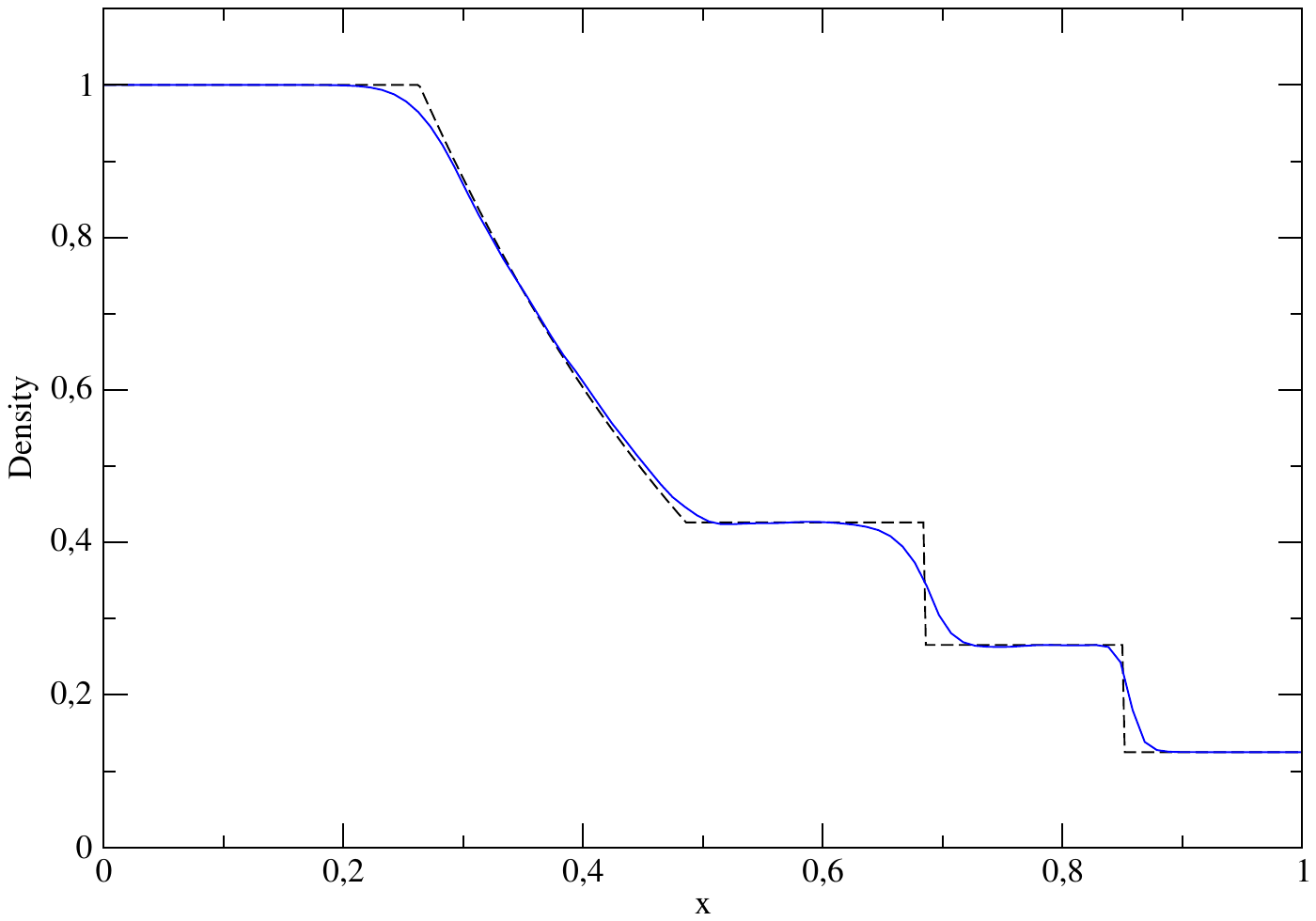}}
\subfigure[Density]{\includegraphics[width=0.48\textwidth,trim=30 50 80 80,clip]{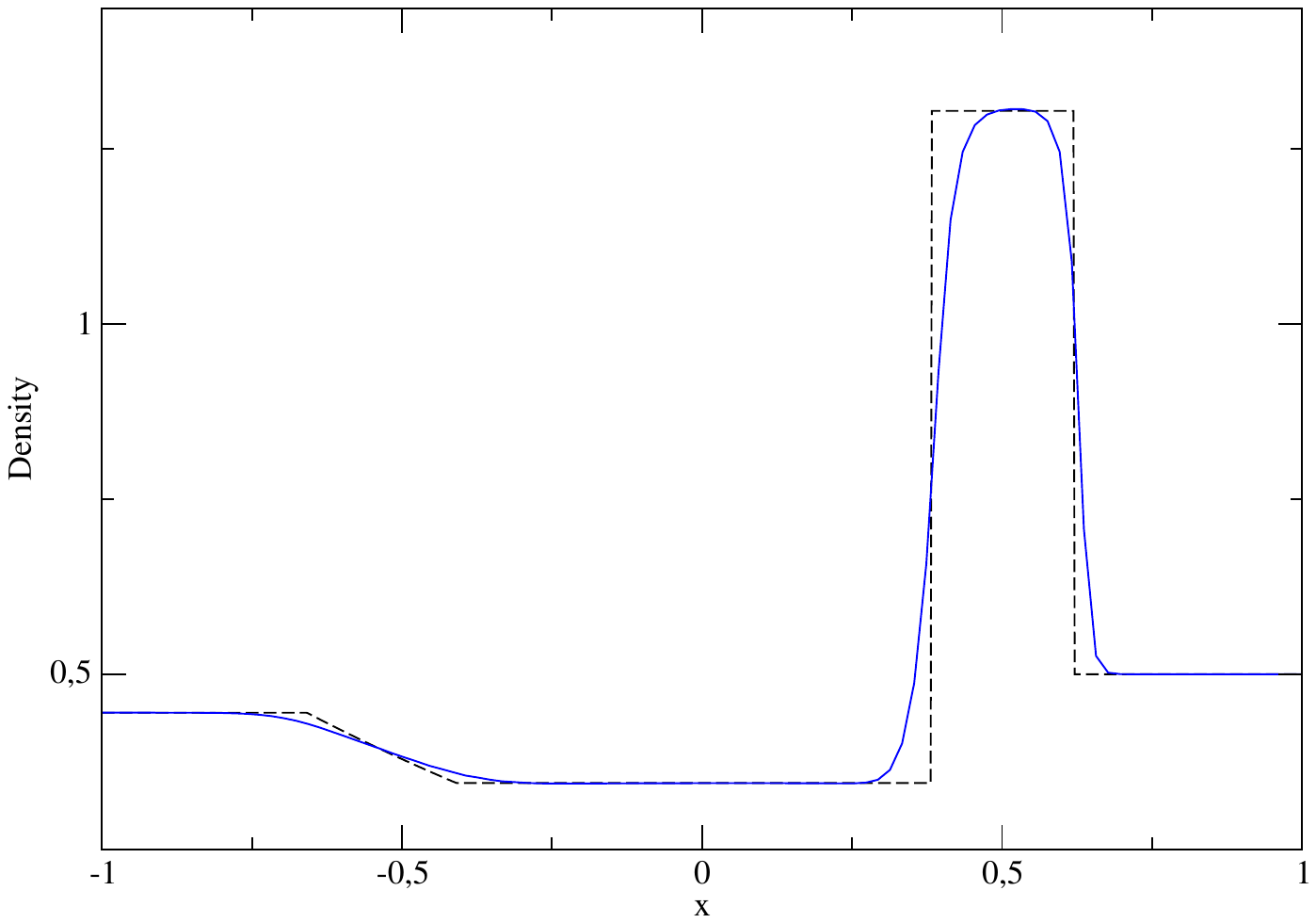}}

\subfigure[Pressure]{\includegraphics[width=0.48\textwidth,trim=30 50 80 80,clip]{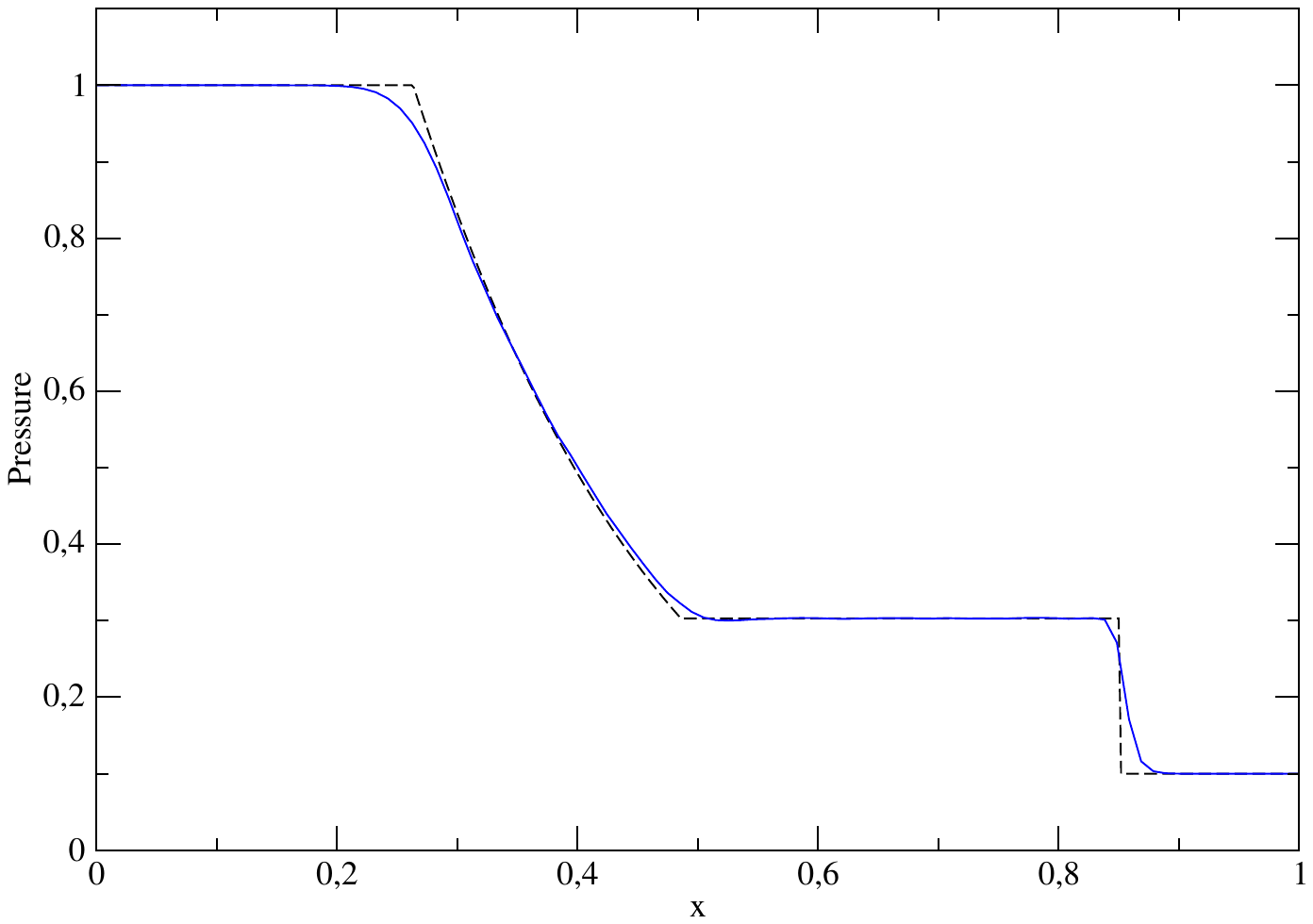}}
\subfigure[Pressure]{\includegraphics[width=0.48\textwidth,trim=30 50 80 80,clip]{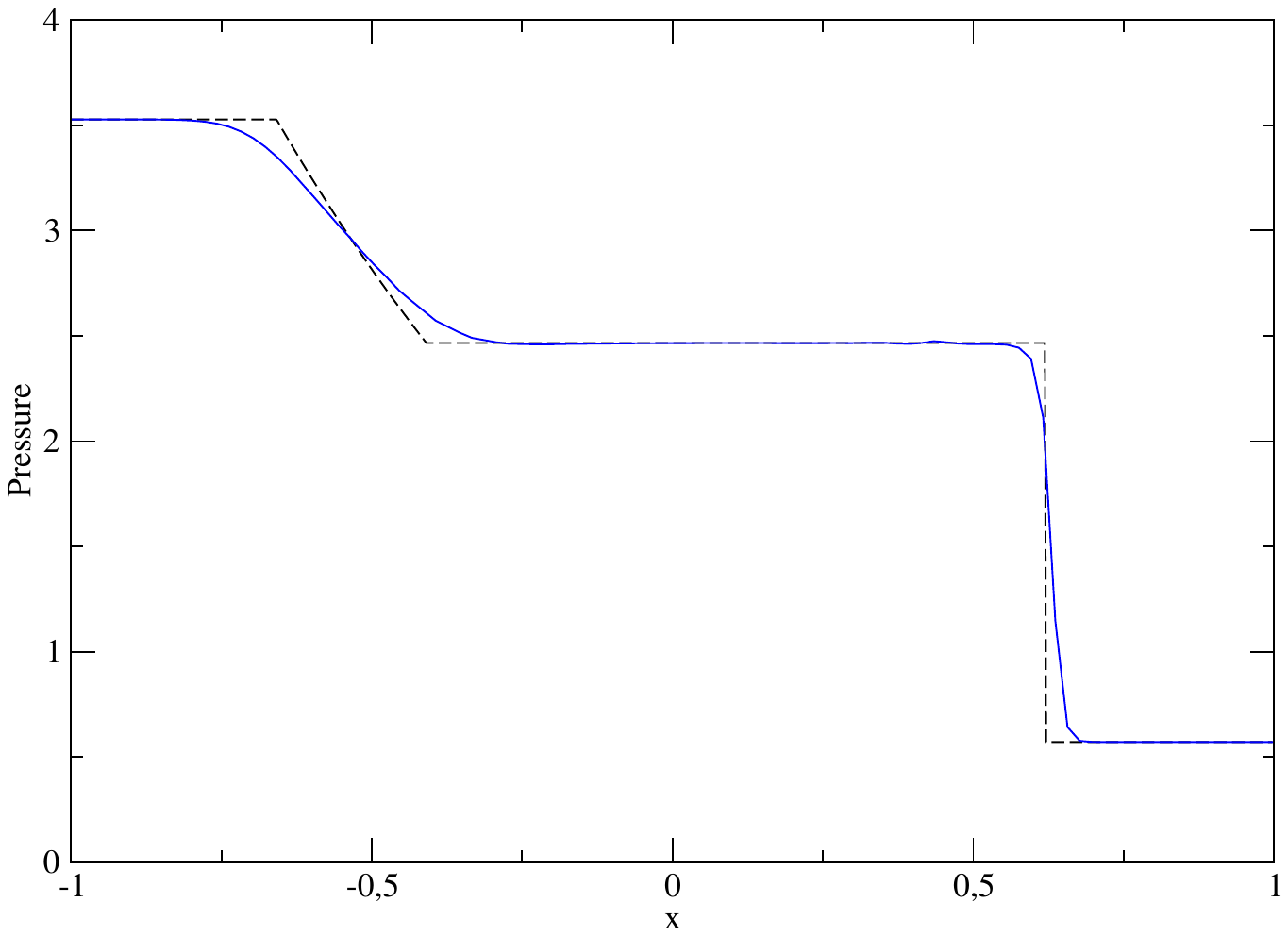}}

\subfigure[Velocity]{\includegraphics[width=0.48\textwidth,trim=30 50 80 80,clip]{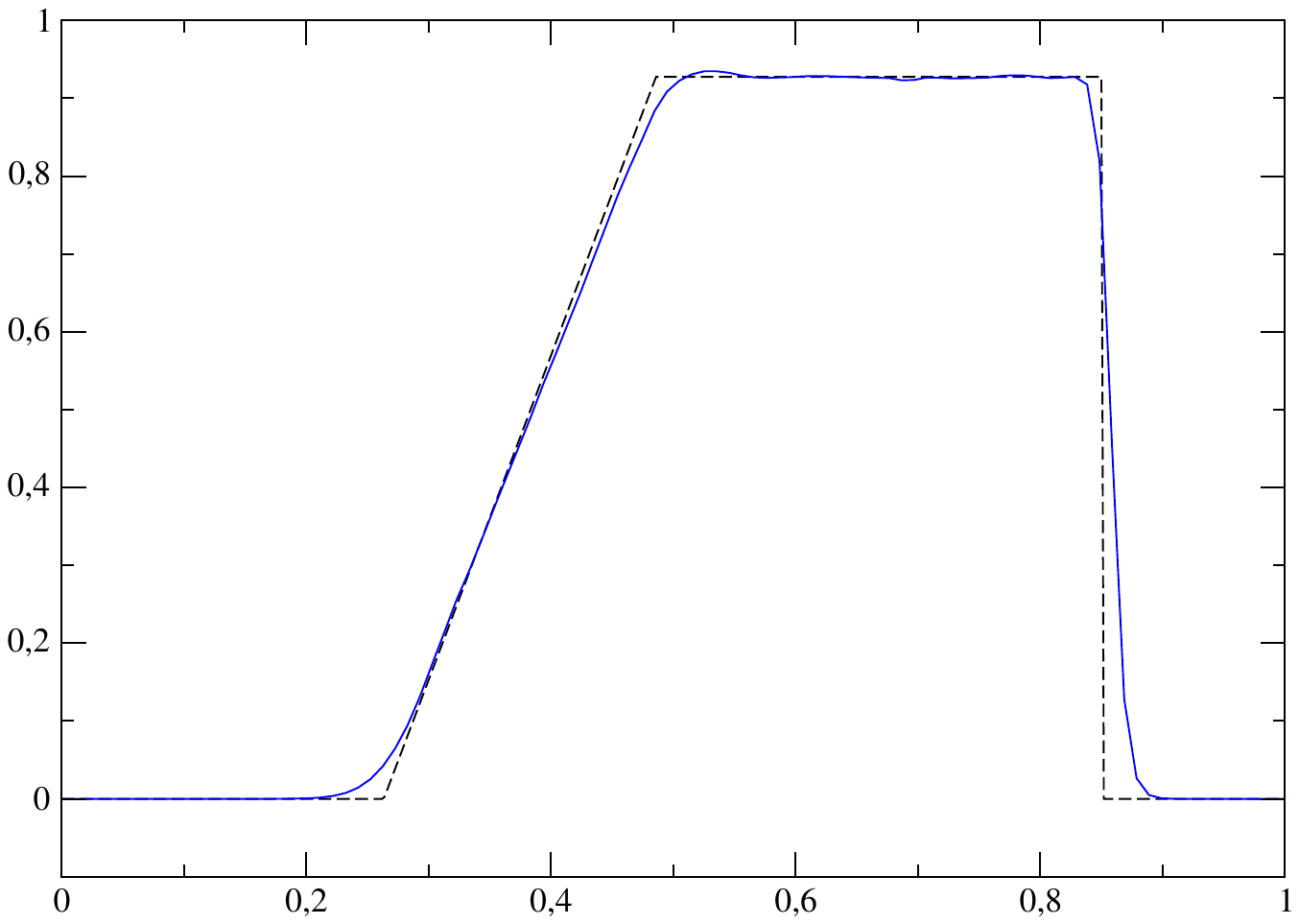}}
\subfigure[Velocity]{\includegraphics[width=0.48\textwidth,trim=30 50 80 80,clip]{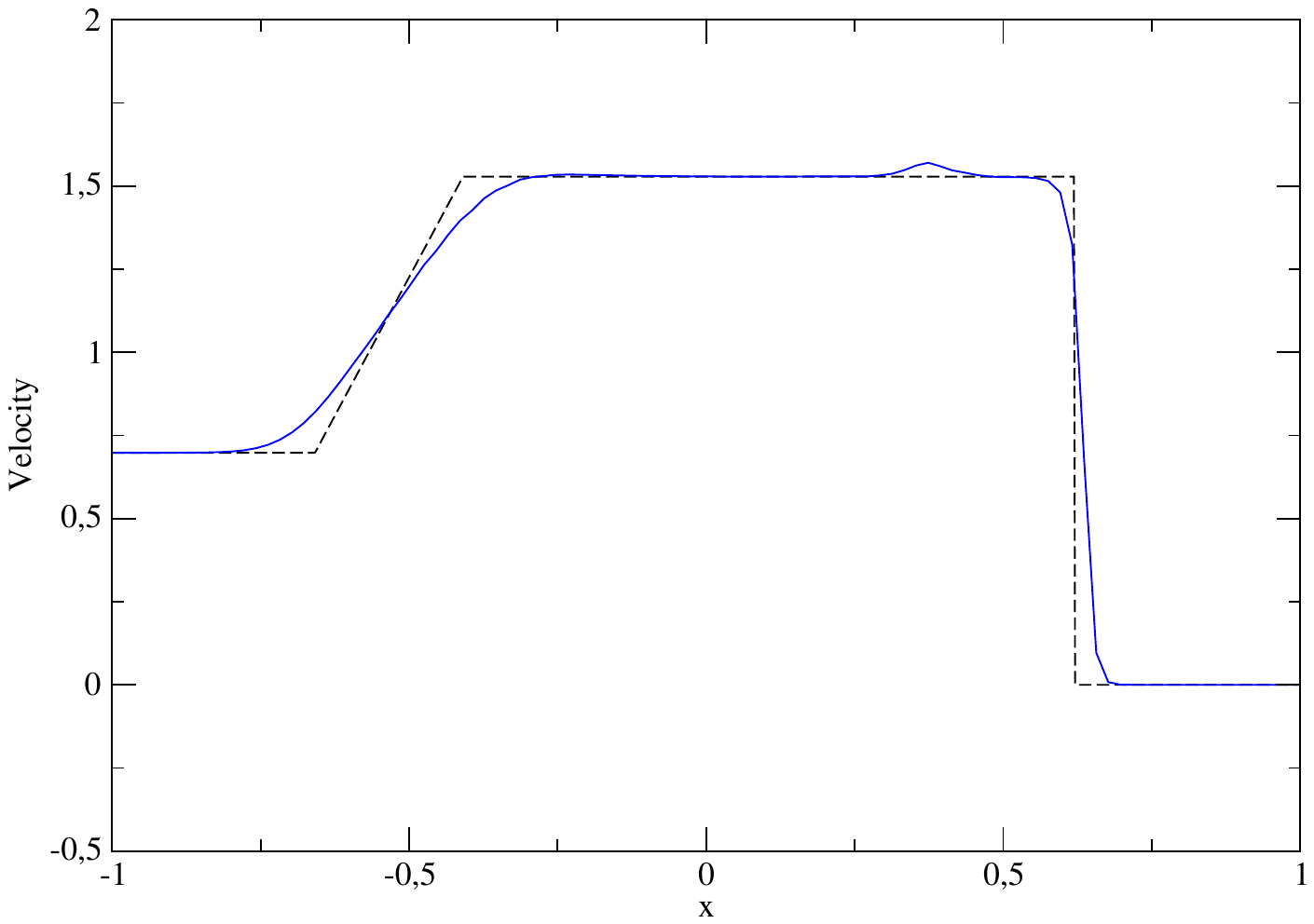}}
\caption{Density, pressure and velocity profiles for the Sod shock tube (left column) and Lax shock tube (right column). The exact solution is displayed in dashed line and the result of the second-order scheme (100 cells) is in solid line.}
\label{fig:sodtac2}
\end{center}
\end{figure}

The results are represented at $t=0.2$s on the left column of fig. \ref{fig:sodtac2} for the density, pressure and velocity computed by the second order scheme. The exact solution is also displayed. The second-order scheme shows a small overall deviation from the reference solution and satisfactory shock velocities are obtained, as expected from a conservative scheme.

We give the error of the solution of the second-order scheme, compared with the exact solution, in table. \ref{table:cvsod}. The calculus of the error is given in equation (\ref{eq:cvtests}). In order to check the spatial convergence only, the time-step is taken as $\Dt = \Dx^2$. We can see that the presence of discontinuities reduces the space convergence order from $2$ to $1$ as in \cite{min2}.

\begin{table}[ht!]
\begin{center}
  \begin{tabular}{|c|c|c|c|}
\hline
Cells & $\Dx$ & $\Dt= \Dx^2$ & $||E||_{L^1}$ \\\hline
100 & 0.01 & $1 \times 10^{-4}$ & $1.3\times 10^{-2}$ \\
200 & 0.005 & $2.5 \times 10^{-5}$ & $6.8\times 10^{-3}$ \\
400 & 0.0025 & $6.25 \times 10^{-6}$ & $3.4\times 10^{-3}$ \\ \hline
\end{tabular}
\caption{Sod shock tube. $L^1$ norm of the relative error between the exact solution and the numerical results for 100, 200 and 400 cells.}
\label{table:cvsod}
\end{center}
\end{table}


\paragraph{Lax shock tube}

This one dimensional shock tube proposed in \cite{lax2005weak} presents stronger shocks than in the Sod shock tube problem. The initial state is divided in a left part ($l$ subscript) for $-1\leq x \leq 0$ and a right part ($r$ subscript) for $0 < x \leq 1$, the initial values being given in table \ref{table:tacLax}
\begin{table}[ht!]
\begin{center}
 \begin{tabular}{|c|c|c|c|}
\hline
      & p  & h  & u  \\ \hline
Left &  3.528   & 27.748 & 0.698      \\ \hline
Right &  0.571  &  3.3997    &  0 \\ \hline
\end{tabular}
\caption{Lax shock tube. Initial conditions for the pressure, the enthalpy and the velocity.}
\label{table:tacLax}
\end{center}
\end{table}\\
We use a perfect gas of equation of state $\rho = \frac{\gamma}{\gamma-1}\frac{p}{h}$, with a constant $\gamma =1.4$, Neumann boundary conditions, and a mesh with 100 cells, with $\Delta x = 0.01$ and $\Dt = 0.001$. 
The Mach number is $\eps = 1$ and the parameter $\alpha$ is zero.

The results are shown at $t=0.25$s on the right column of fig. \ref{fig:sodtac2} for the density, pressure and velocity computed by the second order scheme. The exact solution is also displayed. As in the Sod shock tube, the accuracy is satisfactory and shock velocities are accurately restored.

We give the error of the solution of the second-order scheme, compared with the exact solution, in table. \ref{table:cvlax}. The calculus of the error is given in equation (\ref{eq:cvtests}). We can see that the presence of discontinuities reduces the space convergence order from $2$ to $1$ as in \cite{min2}.
\begin{table}[ht!]
\begin{center}
  \begin{tabular}{|c|c|c|c|}
\hline
Cells & $\Dx$ & $\Dt= \Dx^2$ & $||E||_{L^1}$ \\\hline
100 & 0.02 & $4 \times 10^{-4}$ & $1.2\times 10^{-2}$ \\
200 & 0.04 & $1 \times 10^{-4}$ & $6.1\times 10^{-3}$ \\
400 & 0.005 & $2.5 \times 10^{-5}$ & $3.4\times 10^{-3}$ \\ \hline
\end{tabular}
\caption{Lax shock tube. $L^1$ norm of the relative error between the exact solution and the numerical results for 100, 200 and 400 cells.}
\label{table:cvlax}
\end{center}
\end{table}

\medskip

These test-cases demonstrate the satisfactory behavior of the scheme in the compressible regime.

\subsection{Backward-facing step test-case}

The backward-facing step test-case is a two-dimensional test-case which checks the accuracy of the scheme in the low Mach regime. The geometry of the step creates a region of low velocity where a recirculation of the fluid takes place. The size of the circulation region depends on the Reynolds number of the flow. This test-case has already been treated experimentally and numerically, for example in \cite{armaly1983experimental, zhu1995second}.

This case is modeled by the full Navier-Stokes equations (\ref{eq:NS1})-(\ref{eq:NS3}). The contribution of the diffusive part (right-hand side terms in the equations) is added in explicit source terms as mentioned in part \ref{sec:firstorder}.
\begin{figure}[h!]
\begin{center}
 \includegraphics[width=0.75\textwidth]{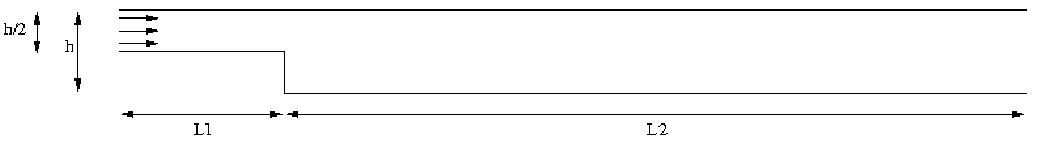}
\caption{Backward-facing step - Geometrical features }
\label{fig:back}
\end{center}
\end{figure} 

The geometry of the step is such that $L_1=4$m, $L_2=18$m, $h=2$m, where the notations refer to fig. \ref{fig:back}. 
The Computational domain is discretized with a uniform Cartesian grid of step $\Delta x=0.2$m, and the time-step is $\Dt = 5\times 10^{-4}$s. The results are displayed at $T=20$s. 

The global Mach number is $\eps=0.01$. We take $\alpha=0$. A perfect gas of equation of state $\rho = \frac{\gamma}{\gamma-1}\frac{p}{h}$ and constant $\gamma=1.4$ is used. The initial conditions are $p=1\times 10^5 \text{ Pa}$, $h=3.5\times 10^4 \text{ J/kg}$, $\uv = (1,0)\text{ m/s}$.	

The coefficients in the diffusive terms of the Navier-Stokes equations are
$ \nu = 1.56\times 10^{-2}$ $m^2/s$, $\lambda = 2.7\times 10^{-2}$ W/m/K, $C_p = \frac{\gamma}{\gamma-1} \frac{R}{M}$ J/K/kg, with $R=8.315$ J/mol/K and $M=0.02897$ kg/mol. The corresponding Reynolds number is $Re\approx 75$.
The external forces are neglected ($\mathbf f_{\text{ext}}=0$).

A wall slip boundary condition ($ \uv \cdot \n = 0 $) is applied on the step and on the top and bottom walls. At the inlet, the velocity and enthalpy are imposed, while a Neumann condition is applied on the pressure. 
The value of the inlet velocity is $\uv=(1,0)$ m/s and the imposed enthalpy is $h=3.5\times 10^4$ J/kg. At the outlet, only the pressure is imposed with a value of $p_{\text{outlet}}=1\times 10^5 \text{ Pa}$.

\medskip

\begin{figure}
\begin{center}
 \subfigure[Second-order Asymptotic Preserving scheme]{\includegraphics[width=0.78\textwidth]{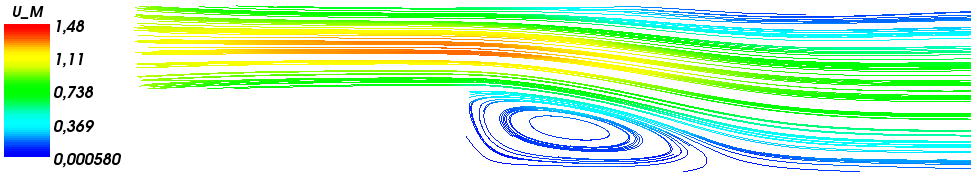}}
 
 \subfigure[Low Mach Roe scheme]{\includegraphics[width=0.78\textwidth]{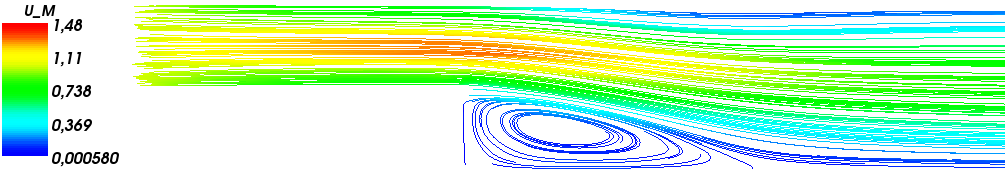}}
 
 \subfigure[Roe scheme]{\includegraphics[width=0.77\textwidth]{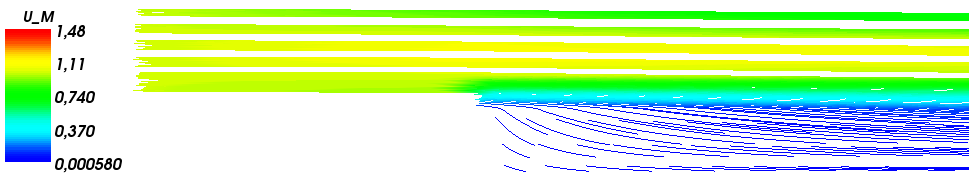}}
\caption{Backward facing step test-case for a Reynolds number $Re=75$ - Streamlines}
\label{fig:backNS_stream75_zoom}
\end{center}
\end{figure}

The modulus of the velocity and the streamlines computed by the second-order Asymptotic Preserving scheme at $t=20s$ are displayed on fig. \ref{fig:backNS_stream75_zoom} for the first $10$m of the channel, whose total length is $22$m. The results are compared to the results obtained with a classical Roe scheme and with the Low Mach Roe scheme mentioned in the introduction of this section.

The second-order Asymptotic Preserving scheme gives a very satisfactory result as the recirculation is computed and matches the dimensions of the recirculation computed by the Low Mach Roe scheme. On the other hand, we can see that the Roe scheme, without a low Mach number treatment, is not able to capture the recirculation of the fluid. This test-case thus confirms that the Asymptotic Preserving scheme has a satisfactory behavior in the incompressible regime.

\subsection{Lid-driven cavity test-case}

The two-dimensional lid-driven cavity test-case is also a well-known problem to assess the adequacy of a code to the low Mach number regime (see for example \cite{rogers1991upwind, ghia1982high}). The case concerns a cubic cavity full of fluid where all the walls are immobile but one : this moving wall drags the neighboring fluid, which initiates a global circulation of the fluid. We expect a central primary recirculation and a smaller lower right eddy.	

This case is also modeled by the full Navier-Stokes equations (\ref{eq:NS1})-(\ref{eq:NS3}), and the contribution of the diffusive part (right-hand side terms in the equations) is added in explicit source terms as mentioned in part \ref{sec:firstorder}.

The global Mach number is $\eps=0.01$. We take $\alpha=0$. A perfect gas of equation of state $\rho = \frac{\gamma}{\gamma-1}\frac{p}{h}$ and constant $\gamma=1.4$ is used. The initial conditions are $p=1\times 10^5 \text{ Pa}$, $ h=3.5\times 10^4 \text{ J/kg}$, $ \uv = (0,0) \text{ m/s}$.

The coefficients in the diffusive terms of the Navier-Stokes equations are
$ \nu = 2.5\times 10^{-2}$ $m^2/s$, $\lambda = 2.7\times 10^{-2}$ W/m/K, $C_p = \frac{\gamma}{\gamma-1} \frac{R}{M}$ J/K/kg, with $R=8.315$ J/mol/K and $M=0.02897$ kg/mol. 
The external forces are neglected ($\mathbf f_{\text{ext}}=0$).

The cavity is formed by the domain $\omega = [0,1]\times[0,1]$, discretized with a uniform cartesian grid of step $\Delta x=1/50$. A wall slip boundary condition ($ \uv \cdot \n = 0 $) is applied on all the walls except the top wall. The top wall is moving at a speed varying continuously from $\uv=0$ at $t=0s$ to $\uv=1$m/s for $t\geq 1s$. The computation has been run with a time-step of $\Dt=2.5\times 10^{-4}$, until a final time of $t=20$s.

\begin{figure}
\begin{center}
 \subfigure[Second-order Asymptotic Preserving scheme]{\includegraphics[width=0.3\textwidth]{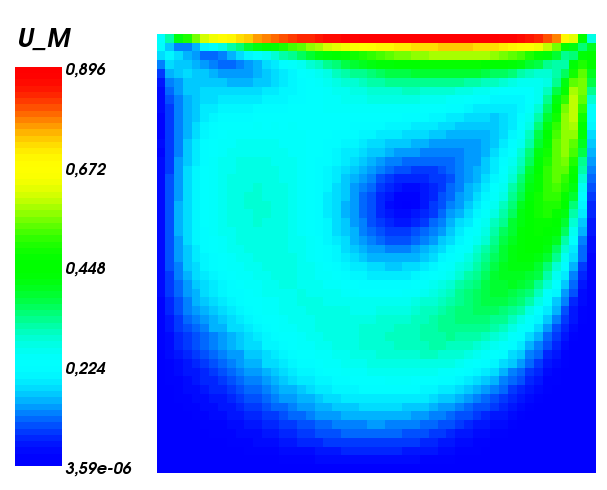}}
 \subfigure[Low Mach Roe scheme ]{\includegraphics[width=0.3\textwidth]{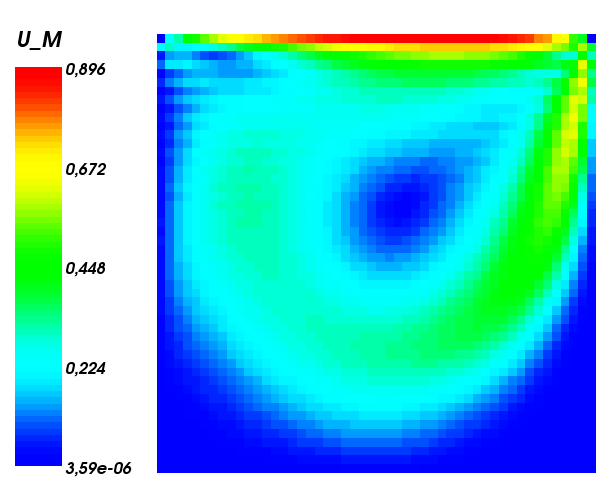}}
 \subfigure[Roe scheme]{\includegraphics[width=0.3\textwidth]{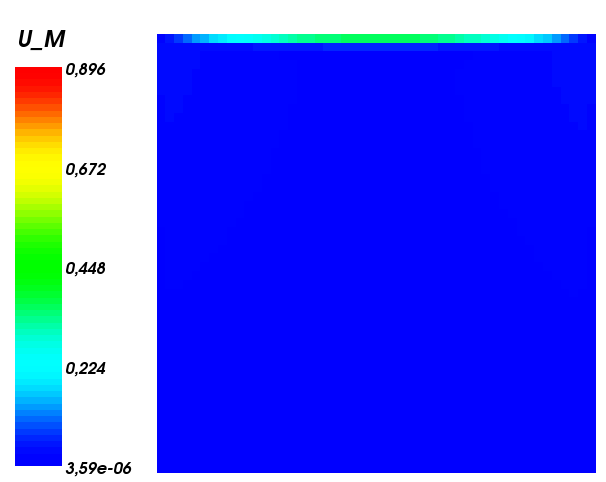}}

 \subfigure[Second-order Asymptotic Preserving scheme]{\includegraphics[width=0.23\textwidth]{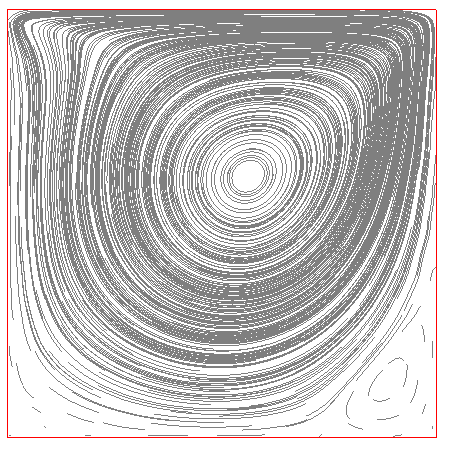}} \hfil
 \subfigure[Low Mach Roe scheme ]{\includegraphics[width=0.23\textwidth]{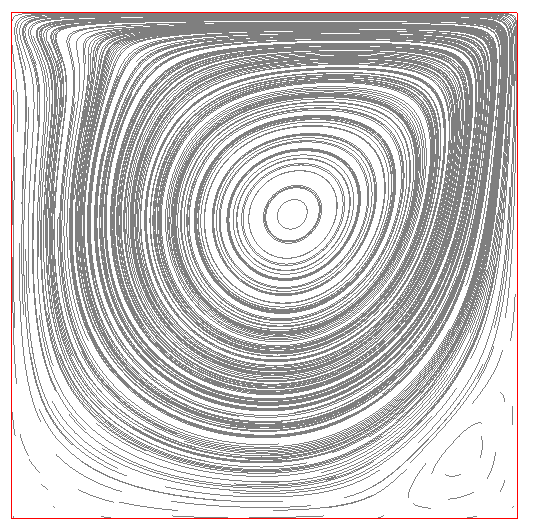}}\hfil
 \subfigure[Roe scheme]{\includegraphics[width=0.23\textwidth]{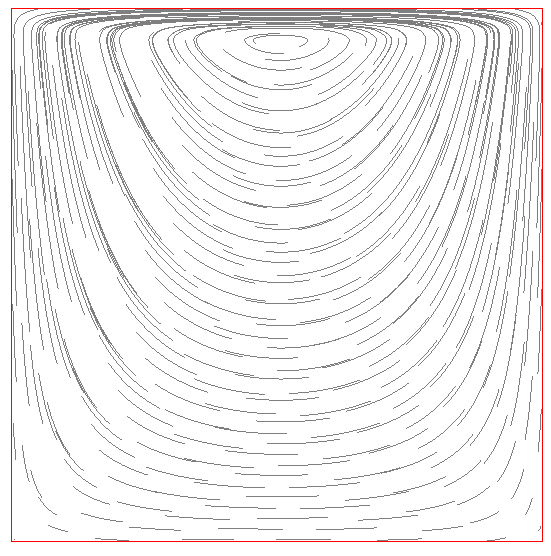}} 

\caption{Lid-driven cavity test-case - Modulus of the velocity and streamlines. }
\label{fig:lid}
\end{center}
\end{figure}

The modulus of the velocity and the streamlines computed by the second-order Asymptotic Preserving scheme at $t=20s$ are displayed fig. \ref{fig:lid}. The results are compared to the results obtained with a classical Roe scheme and with the Low Mach Roe scheme

While the classical Roe scheme displays no recirculation whatsoever, the second-order Asymptotic Preserving scheme shows a good behavior as the circulation region is computed and is similar to the circulation region computed by the Low Mach Roe scheme. The primary vortex is clearly visible for the second-order Asymptotic Preserving scheme and the Low Mach Roe scheme on the figures displaying the streamlines, while it is not correctly computed by the classical Roe scheme. We can also see the lower right eddy expected along with the primary vortex.

\subsection{Heat-driven cavity}

The heat-driven cavity test-case consists in a two-dimensional steady-state single-phase laminar flow resulting from a natural convection created by the difference of temperatures between the two vertical walls of a cubic cavity and the gravity field. The horizontal walls are adiabatic walls. 

This test-case is well suited to evaluate the behavior of a numerical scheme in the low Mach number regime and in the presence of thermal conductivity terms and gravity terms. It has been studied for example in \cite{de1983natural, le2001modeling, dellacherie2010analysis}. This case is very interesting in our situation because it requires the energy equation, contrary to the two previous cases that could have been run with the isentropic Navier-Stokes equations. At last, the global Mach number resulting from the scaling of the equations is $\eps=10^{-4}$, which is much smaller than in the two previous cases. 

This case is modeled by the full Navier-Stokes equations (\ref{eq:NS1})-(\ref{eq:NS3}).A perfect gas of equation of state $\rho = \frac{\gamma}{\gamma-1}\frac{p}{h}$ and constant $\gamma=1.4$ is used.
The dimension of the cubic cavity, $L=1.528\times 10^{-3}$m, is chosen so that the flow is a low Mach number flow ($\eps=10^{-4}$) ; viscosity and conductivity are chosen so that the flow is laminar (low Reynolds number : $Re\approx 37$) and results from natural convection.

The coefficients in the diffusive terms of the Navier-Stokes equations are: $ \nu = 1.619\times 10^{-6}$ $m^2/s$,
$\lambda = 2.29\times 10^{-3}$ W/m/K, $C_p = \frac{\gamma}{\gamma-1} \frac{R}{M}$ J/K/kg, with $R=8.315$ J/mol/K and $M=0.02897$ kg/mol, the external force is gravity: $\mathbf f_{\text{ext}}=(0,-9.81) m/s^2$.

The initial conditions are $p=1\times 10^5 \text{ Pa}$, $ h=2.9167 \times 10^5 \text{ J/kg}$, $\uv = (0,0)\text{ m/s}$.
A wall slip boundary condition ($ \uv \cdot \n = 0 $) is applied on all walls. The velocity of the walls is zero. The top and bottom horizontal walls are adiabatic walls : the thermal flux is imposed to be zero. The temperature is imposed on the right and left vertical walls : $T_l = 283.15$K on the left wall and $T_r=263.15$K on the right wall.

The domain is discretized with a uniform cartesian grid of step $\Delta x=L/40$. The computation has been run with a CFL of $0.002$, until a $t=2s$ and with a convergence criteria for the Newton method of $10^{-7}$. Then the computation has been continued with a CFL of $0.001$ and a convergence criteria of $10^{-10}$ until a final time of $t=3$s. 
This case has been computed with the second-order Asymptotic Preserving scheme, with $\alpha=0$. In order to compare the results, we also present the results of computation of the Roe scheme and of the Low Mach Roe scheme. We expect to find specific patterns in the visualization of the isocontours of the local Mach number and the temperature.

\begin{figure} 
\begin{center}
 \subfigure[Second-order Asymptotic Preserving scheme]{\includegraphics[width=0.3\textwidth]{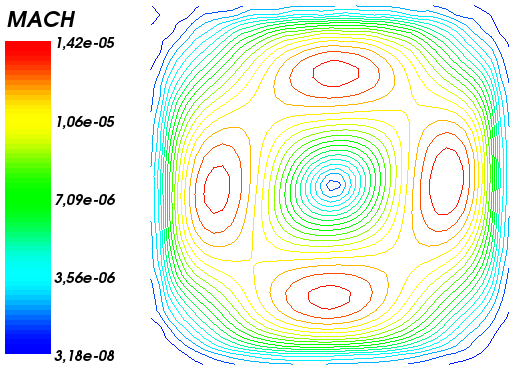}} 
\subfigure[Low Mach Roe scheme ]{\includegraphics[width=0.3\textwidth]{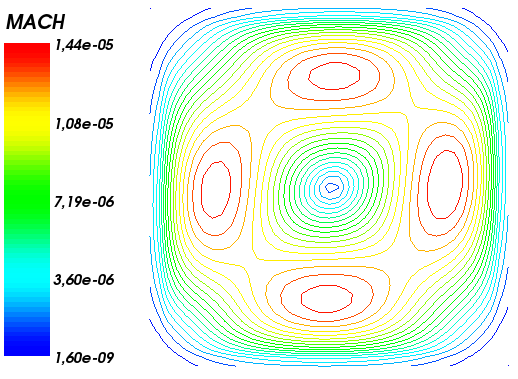}}
\subfigure[Roe scheme ]{\includegraphics[width=0.3\textwidth]{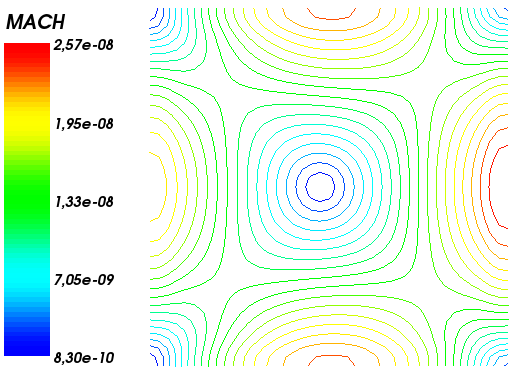}}
\caption{Heat-driven cavity test-case - Isocontours of the Mach number }
\label{fig:heat_mach}
\end{center}
\end{figure}

The isocontours of the local Mach number (which is different from the Mach number $\eps$ resulting from the scaling of the equations) is given on fig. \ref{fig:heat_mach}. We can see that the solution computed by the second-order Asymptotic Preserving scheme matches the solution of the low Mach Roe scheme. On the other hand, the Roe scheme is not able to provide the correct solution and the pattern is very different from the pattern obtained with the scheme adapted to low Mach numbers.
Let us also notice that the local Mach number ranges from $10^{-5}$ to $10^{-9}$, which is very small and confirms that the case lies in the incompressible regime.

\setcounter{equation}{0}
\section{Conclusion}
\label{sec_conclu}

The aim of this paper was to provide an all-speed scheme for the numerical simulation of mixed compressible and incompressible fluid flows. The second-order discretization of the proposed Asymptotic Preserving scheme shows a very good behavior in both flow regimes. In compressible situations, we obtain good shocks properties as the scheme is conservative. In the low Mach number regime, the Asymptotic Preserving property provides a consistent discretization of the incompressible model, the divergence-free condition on the velocity is respected and the pressure is solved via an elliptic equation. The centered spatial discretization of the implicit pressure term allows the time-step to be based on the fluid velocity and not on the acoustic velocity. The time-step can be much larger than with an explicit upwind method and does not depend on the Mach number. The proposed scheme therefore shows a very good behavior on the weakly compressible numerical test-cases such as the backward-facing step and the lid-driven cavity as it provides the expected recirculations of the fluid, and also provides the correct solution on the heat-driven cavity which uses the energy equation. 

Low Mach number regimes are often encountered in multiphase mixtures. The Navier-Stokes equations have been used in this paper as they are very similar to the simplest two-phase flow model, the homogeneous equilibrium model. In future works, we intend to extend the scheme to more elaborate two-phase flow models as the four-equation mixture model and the six-equation two-fluid model. 

 First tests have been realized so far with the four-equation mixture model and a test-case of a water flow in a heated channel has been computed. It has confirmed the ability of the scheme to compute a two-phase mixture, to use a general equation of state (Water and Steam EOS), and to work with heat transfer terms and phase change phenomena.



\end{document}